\documentclass[12pt,twoside, a4paper]{article}
\def\pd{\partial}
\def\mc{\mathcal}

\usepackage[dvips]{graphicx}
\usepackage{amssymb}
\usepackage{amssymb,amsmath}
\usepackage{graphicx}
\usepackage{dsfont}
\usepackage{caption}
\usepackage{subcaption}
\input{epsf.sty}  
\begin{document}
\begin{center}
\LARGE{\textbf{Supersymmetric RG flows and Janus from type II orbifold compactification}}
\end{center}
\vspace{1 cm}
\begin{center}
\large{\textbf{Parinya Karndumri}$^a$ and \textbf{Khem
Upathambhakul}$^b$}
\end{center}
\begin{center}
String Theory and Supergravity Group, Department
of Physics, Faculty of Science, Chulalongkorn University, 254 Phayathai Road, Pathumwan, Bangkok 10330, Thailand
\end{center}
E-mail: $^a$parinya.ka@hotmail.com \\
E-mail: $^b$keima.tham@gmail.com \vspace{1 cm}\\
\begin{abstract}
We study holographic RG flow solutions within four-dimensional $N=4$ gauged supergravity obtained from type IIA and IIB string theories compactified on $T^6/\mathbb{Z}_2\times \mathbb{Z}_2$ orbifold with gauge, geometric and non-geometric fluxes. In type IIB non-geometric compactifications, the resulting gauged supergravity has $ISO(3)\times ISO(3)$ gauge group and admits an $N=4$ $AdS_4$ vacuum dual to an $N=4$ superconformal field theory (SCFT) in three dimensions. We study various supersymmetric RG flows from this $N=4$ SCFT to $N=4$ and $N=1$ non-conformal field theories in the IR. The flows preserving $N=4$ supersymmetry are driven by relevant operators of dimensions $\Delta =1,2$ or alternatively by one of these relevant operators, dual to the dilaton, and irrelevant operators of dimensions $\Delta=4$ while the $N=1$ flows in addition involve marginal deformations. Most of the flows can be obtained analytically. We also give examples of supersymmetric Janus solutions preserving $N=4$ and $N=1$ supersymmetries. These solutions should describe two-dimensional conformal defects within the dual $N=4$ SCFT. Geometric compactifications of type IIA theory give rise to $N=4$ gauged supergravity with $ISO(3)\ltimes U(1)^6$ gauge group. In this case, the resulting gauged supergravity admits an $N=1$ $AdS_4$ vacuum. We also numerically study possible $N=1$ RG flows to non-conformal field theories in this case.
\end{abstract}
\newpage
%%%%%%%%%%%%%%%%%%%%%%%%%%%%%%%%%%%%%%%%%%%%%%%%%%%%%%%%%%%%%%%%%%%%%%%%%%%%%%%%%%%%%%%%%%%%%%%%%%%%%%%%%%%%%%%%%%%%%%%%%%%%%%%%%%%%%%%%%
\section{Introduction}
Along the line of research within the context of the AdS/CFT correspondence, the study of holographic RG flows is of particular interest since the original proposal in \cite{maldacena}. There are many works exploring this type of holographic solutions in various space-time dimensions with different numbers of supersymmetry. In this paper, we will particularly consider holographic RG flows within three-dimensional superconformal field theories (SCFTs) using gauged supergravity in four dimensions. This might give some insight to the dynamics of strongly coupled SCFTs in three dimensions and related brane configurations in string/M-theory.
\\
\indent Most of the previously studied holographic RG flows have been found within the maximal $N=8$ gauged supergravities \cite{Ahn_4D_flow,Flow_in_N8_4D,4D_G2_flow,Warner_M2_flow,Warner_Fisch,Guarino_BPS_DW,Elec_mag_flows,Yi_4D_flow}. Many of these solutions describe various deformations of the $N=8$ SCFTs arising from M2-brane world-volume proposed in \cite{BL,ABJM}. Similar study in the case of lower number of supersymmetry is however less known. For example, a number of RG flow solutions have appeared only recently in $N=3$ and $N=4$ gauged supergravities \cite{N3_SU2_SU3,N3_4D_gauging,tri-sasakian-flow}. In this work, we will give more solutions of this type from the half-maximal $N=4$ gauged supergravity.
\\
\indent $N=4$ supergravity allows for coupling to an arbitrary
number of vector multiplets. With $n$ vector multiplets, there are
$6n+2$ scalars, $2$ from gravity and $6n$ from vector multiplets,
parametrized by $SL(2,\mathbb{R})/SO(2)\times SO(6,n)/SO(6)\times
SO(n)$ coset. The $N=4$ gauged supergravity has been constructed for a
long time in \cite{Eric_N4_4D} and \cite{de_Roo_N4_4D}. Gaugings
constructed in \cite{Eric_N4_4D} are called electric gaugings since
only electric $n+6$ vector fields appearing in the ungauged
Lagrangian gauge a subgroup of $SO(6,n)$. These vector fields transform as a fundamental representation of $SO(6,n)$. The scalar potential of
the resulting gauged supergravity constructed in this way does not
possess any $AdS_4$ critical points \cite{Jan_electricN4,N4_Wagemans}. This
is not the case for the construction in \cite{de_Roo_N4_4D} in which
non-trivial $SL(2,\mathbb{R})$ phases have been included.
\\
\indent The most general gauging in which both electric vector
fields and their magnetic dual can participate has been constructed
in \cite{N4_gauged_SUGRA} using the embedding tensor formalism. A
general gauge group is a subgroup of the full duality group
$SL(2,\mathbb{R})\times SO(6,n)$ with the vector fields and their magnetic dual transforming as a doublet of $SL(2,\mathbb{R})$. In this work, we will consider
$N=4$ gauged supergravity obtained from compactifications of type II
string theories with various fluxes given in
\cite{type_II_orbifold}, for other works along this line see for example \cite{IIA_flux_compact,IIA_flux_stabilization,Dallagata_IIA_flux}.
\\
\indent In \cite{type_II_orbifold}, the scalar potential arising
from flux compactifications of type IIA and IIB theories on
$T^6/\mathbb{Z}_2\times \mathbb{Z}_2$ within a truncation to $SO(3)$
singlet scalars has been considered, and some $AdS_4$ critical
points together with their properties have been given. For type IIB
non-geometric compactification, the vacuum structure is very rich even
with only a few number of fluxes turned on. Among these vacua, there
exists an $N=4$ $AdS_4$ critical point dual to an $N=4$ SCFT with $SO(3)\times SO(3)$ global symmetry. In addition, the full classification of vacua from type IIA geometric compactification has also been
given. In this case, there exist a number of stable
non-supersymmetric $AdS_4$ critical points as well as an $N=1$
$AdS_4$ vacuum, see \cite{N1_AdS4_mIIA} for an $N=1$ supersymmetric $AdS_4$ vacuum in massive type IIA theory.
\\
\indent We are particularly interested in $N=4$ and $N=1$ $AdS_4$ critical points
from these two compactifications. They correspond to $N=4$ and $N=1$ SCFTs in three dimensions with global symmetries $SO(3)\times SO(3)$ and $SO(3)$, respectively. We will look for possible supersymmetric deformations within these two SCFTs in the form of RG flows to non-conformal phases preserving some amount of supersymmetry. These deformations are described by supersymmetric domain walls in the $N=4$ gauged supergravity. In the case of $N=1$ SCFT arising from  massive type IIA theory, non-supersymmetric RG flows to conformal fixed points in the IR have been recently found in \cite{N1_mIIA_flows}.
\\
\indent For type IIB compactification, we will also consider supersymmetric Janus solutions describing $(1+1)$-dimensional conformal interfaces in the $N=4$ SCFT. This type of solutions breaks conformal symmetry in three dimensions but preserves a smaller conformal symmetry on the lower-dimensional interface. Similar to the RG flow solutions, there are only a few examples of these solutions within the context of four-dimensional gauged supergravities \cite{tri-sasakian-flow,warner_Janus,N3_Janus,N4-Janus-Henning-Mario}, see also \cite{Bak_Janus,5D_Janus_CK,5D_Janus_Suh} for examples of higher-dimensional solutions. They also play an important role in the holographic study of interface and boundary CFTs \cite{ICFT_BCFT,dCFT}. We will give more examples of these solutions in $N=4$ gauged supergravity obtained from non-geometric flux compactification.
\\
\indent The paper is organized as follow. In section \ref{N4_SUGRA},
we review relevant formulae and introduce some notations for
$N=4$ gauged supergravity in the embedding tensor formalism. In section \ref{IIB_flow} and \ref{IIB_Janus}, we give a detailed analysis of supersymmetric RG flow and Janus solutions obtained from non-geometric type IIB compactification. Similar study of RG flows from geometric type IIA compactification will be given in section \ref{IIA_flow}. We finally give some conclusions and comments on the results in section \ref{conclusion}. We have also included an appendix containing more details on the conventions and the explicit form of complicated equations.

%%%%%%%%%%%%%%%%%%%%%%%%%%%%%%%%%%%%%%%%%%%%%%%%%%%%%%%%%%%%%%%%%%%%%%%%%%%%%%%%%%%%%%%%%%%%%%%%%%%%%%%%%%%%%%%%%%%%%%%%%%%%%%%%%%%%%%%%%
\section{$N=4$ gauged supergravity coupled to six vector
multiplets}\label{N4_SUGRA} We first review relevant
information and necessary formulae of four-dimensional $N=4$ gauged
supergravity which is the framework we use to find supersymmetric
solutions. We mainly follow the most general gauging of $N=4$
supergravity in the embedding tensor formalism given in
\cite{N4_gauged_SUGRA} in which more details on the construction can
be found. $N=4$ supersymmetry allows for coupling the supergravity
multiplet to an arbitrary number of vector multiplets. We will begin
with a general formulation of $N=4$ gauged supergravity with $n$
vector multiplets and later specify to the case of six vector
multiplets.
\\
\indent In half-maximal $N=4$ supergravity, the supergravity multiplet consists of the graviton
$e^{\hat{\mu}}_\mu$, four gravitini $\psi^i_\mu$, six vectors
$A_\mu^m$, four spin-$\frac{1}{2}$ fields $\chi^i$ and one complex
scalar $\tau$ consisting of the dilaton $\phi$ and the axion $\chi$. The complex scalar can be parametrized by $SL(2,\mathbb{R})/SO(2)$ coset. The supergravity multiplet can couple to an arbitrary number $n$ of vector multiplets, and each vector multiplet contains a vector
field $A_\mu$, four gaugini $\lambda^i$ and six scalars $\phi^m$.
Similar to the dilaton and the axion in the gravity multiplet, the $6n$ scalar fields in these vector multiplets can be parametrized by $SO(6,n)/SO(6)\times SO(n)$ coset.
\\
\indent We will use the following convention on various indices appearing throughout the paper. Space-time and tangent space indices are denoted respectively by $\mu,\nu,\ldots =0,1,2,3$ and
$\hat{\mu},\hat{\nu},\ldots=0,1,2,3$. The $SO(6)\sim SU(4)$
R-symmetry indices will be described by $m,n=1,\ldots, 6$ for the
$SO(6)$ vector representation and $i,j=1,2,3,4$ for the $SO(6)$
spinor or $SU(4)$ fundamental representation. The $n$ vector
multiplets will be labeled by indices $a,b=1,\ldots, n$. All fields in the vector multiplets then carry an additional
index in the form of $(A^a_\mu,\lambda^{ia},\phi^{ma})$. Fermionic fields and the supersymmetry parameters transform in the fundamental representation of $SU(4)_R\sim SO(6)_R$ R-symmetry and are subject to the chirality projections
\begin{equation}
\gamma_5\psi^i_\mu=\psi^i_\mu,\qquad \gamma_5\chi^i=-\chi^i,\qquad \gamma_5\lambda^i=\lambda^i\, .
\end{equation}
On the other hand, for the fields transforming in the anti-fundamental representation of $SU(4)_R$, we have
\begin{equation}
\gamma_5\psi_{\mu i}=-\psi_{\mu i},\qquad \gamma_5\chi_i=\chi_i,\qquad \gamma_5\lambda_i=-\lambda_i\, .
\end{equation}
\indent Gaugings of the matter-coupled $N=4$ supergravity can be
efficiently described by using the embedding tensor. This
tensor encodes all the information about the embedding of any
gauge group $G_0$ in the global or duality symmetry group
$G=SL(2,\mathbb{R})\times SO(6,n)$ in a $G$ covariant way. According to the analysis in \cite{N4_gauged_SUGRA}, a general gauging can be described by two components of the embedding tensor $\xi^{\alpha M}$ and $f_{\alpha MNP}$ with
$\alpha=(+,-)$ and $M,N=(m,a)=1,\ldots, n+6$ denoting fundamental
representations of $SL(2,\mathbb{R})$ and $SO(6,n)$, respectively.
The electric vector fields $A^{+M}=(A^m_\mu,A^a_\mu)$, appearing in
the ungauged Lagrangian, and their magnetic dual $A^{-M}$ form a
doublet under $SL(2,\mathbb{R})$ denoted by $A^{\alpha M}$. A particular electric-magnetic frame in which the $SO(2)\times SO(6,n)$ symmetry, with $SO(2)\subset SL(2,\mathbb{R})$, is manifest in the action can always be chosen. In this frame, $A^{+M}$ and $A^{-M}$ have charges $+1$ and $-1$ under this $SO(2)$.
\\
\indent In general, a subgroup of both $SL(2,\mathbb{R})$ and
$SO(6,n)$ can be gauged, and the magnetic vector fields can also
participate in the gauging. Furthermore, it has been shown in \cite{Jan_electricN4},
see also \cite{N4_Wagemans}, that purely electric gaugings do not admit $AdS_4$
vacua unless an $SL(2,\mathbb{R})$ phase is included \cite{de_Roo_N4_4D}. The latter is however incorporated in the magnetic component $f_{-MNP}$ \cite{N4_gauged_SUGRA}. Accordingly, we will consider only gaugings involving both
electric and magnetic vector fields in order to obtain $AdS_4$ vacua. We will see that gauged supergravities obtained from type II compactifications are precisely of this form.
\\
\indent The gauge covariant derivative can be written as
\begin{equation}
D_\mu=\nabla_\mu-gA_\mu^{\alpha M}\Theta_{\alpha M}^{\phantom{\alpha
M}NP}t_{NP}+gA_\mu^{M(\alpha}\epsilon^{\beta)\gamma}\xi_{\gamma
M}t_{\alpha\beta}
\end{equation}
where $\nabla_\mu$ is the usual space-time covariant derivative including the spin connection.
$t_{MN}$ and $t_{\alpha\beta}$ are $SO(6,n)$ and $SL(2,\mathbb{R})$
generators which can be chosen as
\begin{equation}
(t_{MN})_P^{\phantom{P}Q}=2\delta^Q_{[M}\eta_{N]P},\qquad
(t_{\alpha\beta})_\gamma^{\phantom{\gamma}\delta}=2\delta^\delta_{(\alpha}\epsilon_{\beta)\gamma}
\end{equation}
with $\epsilon^{\alpha\beta}=-\epsilon^{\beta\alpha}$ and
$\epsilon^{+-}=1$.
$\eta_{MN}=\textrm{diag}(-1,-1,-1,-1,-1,-1,1,\ldots,1)$ is the
$SO(6,n)$ invariant tensor, and $g$ is the gauge coupling constant
that can be absorbed in the embedding tensor $\Theta$.
\\
\indent The embedding tensor component $\Theta_{\alpha MNP}$ can be written in terms of
$\xi^{\alpha M}$ and $f_{\alpha MNP}$ components as
\begin{equation}
\Theta_{\alpha MNP}=f_{\alpha MNP}-\xi_{\alpha[N}\eta_{P]M}\, .
\end{equation}
To define a consistent gauging, the embedding tensor has to satisfy a quadratic constraint. This ensures that the gauge generators
\begin{equation}
X_{\alpha M}=\Theta_{\alpha MNP}t^{NP}-\xi^\beta_M t_{\alpha\beta}
\end{equation}
form a closed algebra.
\\
\indent In this work, we will consider solutions with
only the metric and scalars non-vanishing. In addition, we will consider gaugings with only $f_{\alpha MNP}$ non-vanishing. Therefore, we will set all vector fields and $\xi_{\alpha M}$ to zero from now on. In particular, this simplifies the full quadratic constraint to
\begin{equation}
f_{\alpha R[MN}f_{\beta PQ]}^{\phantom{\beta PQ}R}=0,\qquad \epsilon^{\alpha\beta}f_{\alpha MNR}f_{\beta PQ}^{\phantom{\beta PQ}R}=0\, .
\end{equation}
For electric gaugings, these relations reduce to the usual Jacobi identity for $f_{MNP}=f_{+MNP}$ as shown in \cite{Eric_N4_4D} and \cite{de_Roo_N4_4D}.
\\
\indent The scalar coset manifold $SL(2,\mathbb{R})/SO(2)\times SO(6,n)/SO(6)\times SO(n)$ can be described by the coset representative $(\mc{V}_\alpha,\mc{V}_M^{\phantom{M}A})$ with $A=(m,a)$. The first factor can be parametrized by
\begin{equation}
\mc{V}_\alpha=\frac{1}{\sqrt{\textrm{Im} \tau}}\left(
                                         \begin{array}{c}
                                           \tau \\
                                           1 \\
                                         \end{array}
                                       \right)
\end{equation}
or equivalently by a symmetric $2\times 2$ matrix
\begin{equation}
M_{\alpha\beta}=\textrm{Re} (\mc{V}_\alpha\mc{V}^*_\beta)=\frac{1}{\textrm{Im}
\tau}\left(
                                    \begin{array}{cc}
                                      |\tau|^2 & \textrm{Re} \tau \\
                                      \textrm{Re} \tau & 1 \\
                                    \end{array}
                                  \right).
\end{equation}
Note also that $\textrm{Im}(\mc{V}_\alpha\mc{V}^*_\beta)=\epsilon_{\alpha\beta}$.
The complex scalar $\tau$ can also be written in terms of the
dilaton $\phi$ and the axion $\chi$ as
\begin{equation}
\tau=\chi+ie^\phi\, .
\end{equation}
\indent For the $SO(6,n)/SO(6)\times SO(n)$ factor, we introduce another
coset representative $\mc{V}_M^{\phantom{M}A}$ transforming by
left and right multiplications under $SO(6,n)$ and $SO(6)\times
SO(n)$, respectively. From the splitting of the index $A=(m,a)$, we can write the coset representative as
$\mc{V}_M^{\phantom{M}A}=(\mc{V}_M^{\phantom{M}m},\mc{V}_M^{\phantom{M}a})$.
Being an element of $SO(6,n)$, the matrix $\mc{V}_M^{\phantom{M}A}$
satisfies the relation
\begin{equation}
\eta_{MN}=-\mc{V}_M^{\phantom{M}m}\mc{V}_N^{\phantom{M}m}+\mc{V}_M^{\phantom{M}a}\mc{V}_N^{\phantom{M}a}\,
.
\end{equation}
As in the $SL(2,\mathbb{R})/SO(2)$ factor, we can parametrize the
$SO(6,n)/SO(6)\times SO(n)$ coset in term of a symmetric matrix
\begin{equation}
M_{MN}=\mc{V}_M^{\phantom{M}m}\mc{V}_N^{\phantom{M}m}+\mc{V}_M^{\phantom{M}a}\mc{V}_N^{\phantom{M}a}\,
.
\end{equation}
\indent We are now in a position to give the bosonic Lagrangian with
the vector fields and auxiliary two-form fields vanishing
\begin{equation}
e^{-1}\mc{L}=\frac{1}{2}R+\frac{1}{16}\pd_\mu M_{MN}\pd^\mu
M^{MN}-\frac{1}{4(\textrm{Im}\tau)^2}\pd_\mu \tau \pd^\mu \tau^*-V
\end{equation}
where $e$ is the vielbein determinant. The scalar potential can be written in terms of scalar coset representatives and the embedding tensor as
\begin{eqnarray}
V&=&\frac{g^2}{16}\left[f_{\alpha MNP}f_{\beta
QRS}M^{\alpha\beta}\left[\frac{1}{3}M^{MQ}M^{NR}M^{PS}+\left(\frac{2}{3}\eta^{MQ}
-M^{MQ}\right)\eta^{NR}\eta^{PS}\right]\right.\nonumber \\
& &\left.-\frac{4}{9}f_{\alpha MNP}f_{\beta
QRS}\epsilon^{\alpha\beta}M^{MNPQRS}\right]
\end{eqnarray}
where $M^{MN}$ is the inverse of $M_{MN}$, and $M^{MNPQRS}$ is
defined by
\begin{equation}
M_{MNPQRS}=\epsilon_{mnpqrs}\mc{V}_{M}^{\phantom{M}m}\mc{V}_{N}^{\phantom{M}n}
\mc{V}_{P}^{\phantom{M}p}\mc{V}_{Q}^{\phantom{M}q}\mc{V}_{R}^{\phantom{M}r}\mc{V}_{S}^{\phantom{M}s}\label{M_6}
\end{equation}
with indices raised by $\eta^{MN}$.
\\
\indent Before giving an explicit parametrization of the scalar coset, we give fermionic supersymmetry transformations of $N=4$ gauged supergravity which play an important role in subsequent analyses. These are given by
\begin{eqnarray}
\delta\psi^i_\mu &=&2D_\mu \epsilon^i-\frac{2}{3}gA^{ij}_1\gamma_\mu
\epsilon_j,\\
\delta \chi^i &=&i\epsilon^{\alpha\beta}\mc{V}_\alpha D_\mu
\mc{V}_\beta\gamma^\mu \epsilon^i-\frac{4}{3}igA_2^{ij}\epsilon_j,\\
\delta \lambda^i_a&=&2i\mc{V}_a^{\phantom{a}M}D_\mu
\mc{V}_M^{\phantom{M}ij}\gamma^\mu\epsilon_j+2igA_{2aj}^{\phantom{2aj}i}\epsilon^j\,
.
\end{eqnarray}
The fermion shift matrices, appearing in fermionic mass-like terms in the gauged Lagrangian, are defined by
\begin{eqnarray}
A_1^{ij}&=&\epsilon^{\alpha\beta}(\mc{V}_\alpha)^*\mc{V}_{kl}^{\phantom{kl}M}\mc{V}_N^{\phantom{N}ik}
\mc{V}_P^{\phantom{P}jl}f_{\beta M}^{\phantom{\beta M}NP},\nonumber
\\
A_2^{ij}&=&\epsilon^{\alpha\beta}\mc{V}_\alpha\mc{V}_{kl}^{\phantom{kl}M}\mc{V}_N^{\phantom{N}ik}
\mc{V}_P^{\phantom{P}jl}f_{\beta M}^{\phantom{\beta M}NP},\nonumber
\\
A_{2ai}^{\phantom{2ai}j}&=&\epsilon^{\alpha\beta}\mc{V}_\alpha
\mc{V}^M_{\phantom{M}a}\mc{V}^N_{\phantom{N}ik}\mc{V}_P^{\phantom{P}jk}f_{\beta
MN}^{\phantom{\beta MN}P}
\end{eqnarray}
where $\mc{V}_M^{\phantom{M}ij}$ is defined in terms of the 't Hooft
symbols $G^{ij}_m$ and $\mc{V}_M^{\phantom{M}m}$ as
\begin{equation}
\mc{V}_M^{\phantom{M}ij}=\frac{1}{2}\mc{V}_M^{\phantom{M}m}G^{ij}_m
\end{equation}
and similarly for its inverse
\begin{equation}
\mc{V}^M_{\phantom{M}ij}=-\frac{1}{2}\mc{V}^M_{\phantom{M}m}(G^{ij}_m)^*\,
.
\end{equation}
$G^{ij}_m$ convert an index $m$ in vector representation of $SO(6)$ to an anti-symmetric pair of indices $[ij]$ in the $SU(4)$ fundamental representation. They satisfy the relations
\begin{equation}
G_{mij}=-(G^{ij}_m)^*=-\frac{1}{2}\epsilon_{ijkl}G^{kl}_m\, .
\end{equation}
The explicit form of these matrices can be found in the appendix.
\\
\indent We finally note the expression for the scalar potential written in terms of $A_1$ and $A_2$ tensors as
\begin{equation}
V=-\frac{1}{3}A^{ij}_1A_{1ij}+\frac{1}{9}A^{ij}_2A_{2ij}+\frac{1}{2}A_{2ai}^{\phantom{2ai}j}
A_{2a\phantom{i}j}^{\phantom{2a}i}\, .
\end{equation}
It follows that unbroken supersymmetry corresponds to an eigenvalue of $A^{ij}_1$, $\alpha$, satisfying $V_0=-\frac{\alpha^2}{3}$ where $V_0$ is the value of the scalar potential at the vacuum, the cosmological constant.
\\
\indent We now consider the case of $n=6$ vector multiplets. Possible gauge groups are accordingly subgroups of $SO(6,6)$ for $\xi_{\alpha M}=0$. Following \cite{type_II_orbifold}, we restrict ourselves to solutions preserving at least $SO(3)$ subgroup of the full gauge group. The residual $SO(3)$ symmetry is embedded in $SO(6,6)$ as a diagonal subgroup of $SO(3)\times SO(3)\times SO(3)\times SO(3)$ with the four factors of $SO(3)$ being subgroups of $SO(6)\times SO(6)\subset SO(6,6)$. The $36$ scalars within $SO(6,6)/SO(6)\times SO(6)$ coset transform as $(\mathbf{6},\mathbf{6})$ under $SO(6)\times SO(6)$ compact subgroup. The above embedding of $SO(3)\times SO(3)$ in $SO(6)$ is given by
\begin{equation}
\mathbf{6}\rightarrow (\mathbf{3},\mathbf{1})+(\mathbf{1},\mathbf{3}).\label{6_rep}
\end{equation}
This implies that the $36$ scalars transform as
\begin{equation}
(\mathbf{6},\mathbf{6})\rightarrow 4\times (\mathbf{1}+\mathbf{3}+\mathbf{5})
\end{equation}
under the unbroken $SO(3)\sim [SO(3)\times SO(3)\times SO(3)\times SO(3)]_{\textrm{diag}}$. We see that there are four $SO(3)$ singlets. We will denote these scalars by $(\varphi_1,\varphi_2,\chi_1,\chi_2)$ as in \cite{type_II_orbifold}. In addition, we will also use the explicit parametrization given in \cite{type_II_orbifold}. This gives the coset representative
\begin{equation}
\mc{V}_M^{\phantom{M}A}=\left(
                          \begin{array}{cc}
                            e^T & 0 \\
                            Be^T & e^{-1} \\
                          \end{array}
                        \right)\otimes \mathbb{I}_3
\end{equation}
where the two $2\times 2$ matrices $e$ and $B$ are defined by
\begin{equation}
e=e^{\frac{1}{2}(\varphi_1+\varphi_2)}\left(
                                        \begin{array}{cc}
                                          1 & \chi_2 \\
                                          0 & e^{-\varphi_2} \\
                                        \end{array}
                                      \right),\qquad
                                      B=\left(
                                          \begin{array}{cc}
                                            0 & \chi_1 \\
                                            -\chi_1 & 0 \\
                                          \end{array}
                                        \right).
\end{equation}
Explicitly, the $SO(6,6)/SO(6)\times SO(6)$ coset representative consisting of all $SO(3)$ singlet scalars is given by
\begin{equation}
{\mathcal{V}_{M}}^{A}=\left(
\begin{array}{cccc}
 e^{\frac{1}{2} \left(\varphi _1+\varphi _2\right)} & 0 & 0 & 0 \\
 e^{\frac{1}{2} \left(\varphi _1+\varphi _2\right)} \chi _2 & e^{\frac{1}{2}
   \left(\varphi _1-\varphi _2\right)} & 0 & 0 \\
 e^{\frac{1}{2} \left(\varphi _1+\varphi _2\right)} \chi _1 \chi _2 & e^{\frac{1}{2}
   \left(\varphi _1-\varphi _2\right)} \chi _1 & e^{-\frac{1}{2} \left(\varphi
   _1+\varphi _2\right)} & -e^{\frac{1}{2} \left(\varphi _2-\varphi _1\right)} \chi _2
   \\
 -e^{\frac{1}{2} \left(\varphi _1+\varphi _2\right)} \chi _1 & 0 & 0 & e^{\frac{1}{2}
   \left(\varphi _2-\varphi _1\right)} \\
\end{array}
\right)
\otimes\mathbb{I}_{3}\, .
\end{equation}
It should also be noted that there are two scalars which are singlet under $SO(3)\times SO(3)\subset [SO(3)\times SO(3)]_{\textrm{diag}}\times [SO(3)\times SO(3)]_{\textrm{diag}}$ as can be seen by taking the tensor product of the representation $\mathbf{6}$ in \eqref{6_rep} giving rise to two singlets $(\mathbf{1},\mathbf{1})$ of $SO(3)\times SO(3)$. These two singlets correspond to $\varphi_1$ and $\varphi_2$.
\\
\indent Similarly, the $SL(2,\mathbb{R})/SO(2)$ coset representative will be parametrized by
\begin{equation}
\mathcal{V}_{\alpha}=e^{\varphi _g/2}\left(
\begin{array}{c}
 \chi _g-i e^{-\varphi _g} \\
 1 \\
\end{array}
\right).
\end{equation}
With all these and the definition $\phi^i=(\varphi_g,\varphi_1,\varphi_2,\chi_g,\chi_1,\chi_2)$, the scalar kinetic terms can be found to be
\begin{eqnarray}
\mathcal{L}_{\text{kin}} &=& -\frac{1}{2}K_{ij}\partial_\mu \phi^{i}\partial^\mu \phi^{j}\nonumber \\
&=& -\frac{1}{4} \left(\pd_\mu \varphi_g\pd^\mu \varphi_g+3\pd_\mu \varphi_1\pd^\mu \varphi_1+3\pd_\mu \varphi_2\pd^\mu \varphi_2+e^{2 \varphi _g} \pd_\mu\chi _g\pd^\mu\chi_g\right.\nonumber \\
& &\left.+3 e^{2 \varphi _1} \pd_\mu\chi _1\pd^\mu\chi_1+3 e^{2 \varphi _2} \pd_\mu\chi _2\pd^\mu\chi_2\right)\label{scalar_kin}
\end{eqnarray}
where we have defined the scalar metric $K_{ij}$ which will play a role in writing the BPS equations.
\\
\indent The four $SO(3)$ singlet scalars in $SO(6,6)/SO(6)\times SO(6)$ correspond to non-compact generators of $SO(2,2)\subset SO(6,6)$ that commute with the $SO(3)$ symmetry. It is convenient to split indices $M=(AI)$ for $A=1,2,3,4$ and $I=1,2,3$. This implies that the $SO(6,6)$ fundamental representation decomposes as $(\mathbf{4},\mathbf{3})$ under $SO(2,2)\times SO(3)$. In terms of $(AI)$ indices, the embedding tensor can be written as
\begin{equation}
f_{\alpha MNP} =f_{\alpha AIBJCK}=\Lambda_{\alpha ABC}\epsilon_{IJK}
\end{equation}
with $\Lambda_{\alpha ABC}=\Lambda_{\alpha (ABC)}$. In particular, the quadratic constraints read
\begin{equation}
\epsilon^{\alpha\beta}{\Lambda_{\alpha AB}}^C\Lambda_{\beta DEC}=0,\qquad {\Lambda_{(\alpha A[B}}^C\Lambda_{\beta)D]EC}=0\, .
\end{equation}
\indent
The $SO(6,6)$ fundamental indices $M,N$ can also be decomposed into $(m,\bar{m})$, $m,\bar{m}=1,2,\ldots, 6$. In connection with the internal manifold $T^6/\mathbb{Z}_2\times \mathbb{Z}_2$, the index $m$ is used to label the $T^6$ coordinates and split into $(a,i)$ such that $a=1,3,5$ and $i=2,4,6$. Similar decomposition is also in use for $\bar{m}=(\bar{a},\bar{i})$. All together, indices $A,B$ can be written as $A=(1,2,3,4)=(a,i,\bar{a},\bar{i})$. Indices $I,J=1,2,3$ label the three $T^2$'s inside $T^6\sim T^2\times T^2\times T^2$.
\\
\indent The $SO(6,6)$ invariant tensor $\eta_{MN}$ and its inverse are chosen to be
\begin{equation}
\eta_{MN}=\eta^{MN}=\left(
                      \begin{array}{cc}
                        0 & \mathbb{I}_6 \\
                        \mathbb{I}_6 & 0 \\
                      \end{array}
                    \right).
\end{equation}
This leads to some extra projections on the negative and positive eigenvalues of $\eta_{MN}$. For example, in order to compute $M_{MNPQRS}$ in the scalar potential defined by equation \eqref{M_6}, we need to project the second index of ${\mc{V}_M}^A$ by using the projection matrix
\begin{equation}
R=\frac{1}{\sqrt{2}}\left(
\begin{array}{cc}
 -\mathbb{I}_{6} & \mathbb{I}_{6} \\
 \mathbb{I}_{6} & \mathbb{I}_{6} \\
\end{array}
\right).
\end{equation}
Finally, we will also set the gauge coupling $g=\frac{1}{2}$ as in \cite{type_II_orbifold}.

%%%%%%%%%%%%%%%%%%%%%%%%%%%%%%%%%%%%%%%%%%%%%%%%%%%%%%%%%%%%%%%%%%%%%%%%%%%%%%%%%%%%%%%%%%%%%%%%%%%%%%%%%%%%%%%%%%%%%%%%%%%%%%%%%%%%%%%%%
\section{RG flows from type IIB non-geometric compactification}\label{IIB_flow}
We begin with a non-geometric compactification of type IIB theory on $T^6/\mathbb{Z}_2\times \mathbb{Z}_2$. This involves the fluxes of NS and RR three-form fields $(H_3,F_3)$ and non-geometric $(P,Q)$ fluxes. This compactification admits a locally geometric description although it is non-geometric in nature.
\\
\indent From the result of \cite{type_II_orbifold}, the effective $N=4$ gauged supergravity theory is not unique. In this paper, we will only consider the gauged supergravity admitting the maximally supersymmetric $N=4$ $AdS_4$ vacuum. In this case, all the gauge and non-geometric fluxes lead to the following components of the embedding tensor
\begin{eqnarray}
f_{-\bar{i}\bar{j}\bar{k}}&=&\Lambda_{-444}=-\lambda,\qquad f_{+\bar{a}\bar{b}\bar{c}}=\Lambda_{+333}=\lambda,\nonumber \\
f_{-\bar{i}\bar{j}k}&=&\Lambda_{-244}=-\lambda,\qquad f_{+a\bar{b}\bar{c}}=\Lambda_{+133}=\lambda
\end{eqnarray}
for a constant $\lambda$. The first and second lines correspond to $(H_3,F_3)$ and $(P,Q)$ fluxes, respectively. As shown in \cite{type_II_orbifold}, the gauge group arising from this embedding tensor is $ISO(3)\times ISO(3)\sim [SO(3)\ltimes T^3]\times[SO(3)\ltimes T^3]$. This gauge group is embedded in $SO(6,6)$ via the $SO(3,3)\times SO(3,3)$ subgroup.
\\
\indent Using this embedding tensor and the explicit form of the scalar coset representative given in the previous section, we find the scalar potential
\begin{align}
V =&\frac{1}{32} e^{\varphi _1-3 \varphi _2-\varphi _g} \lambda ^2 \left[e^{2 \varphi _1}-3 e^{2 \varphi _2}+6 e^{\varphi _1+2 \varphi _2+\varphi _g}-18 e^{3 \varphi _2+\varphi _g}-3 e^{4 \varphi _2+2 \varphi _g}\right.\notag\\
   &-2 e^{2 \varphi _1+3 \varphi _2+\varphi _g} (1+3 \chi _1^2)+3 e^{2   (\varphi _1+\varphi _2)} (\chi _1-\chi _2)^2-12 e^{5 \varphi _2+\varphi _g} \chi _2^2\notag\\
   &+3 e^{6 \varphi _2} \chi _2^4+e^{2   \varphi _1+6 \varphi _2} \chi _2^4 (\chi _2-3 \chi _1)^2+3 e^{2 \varphi _1+4 \varphi _2} \chi _2^2 (\chi  _2-2 \chi _1)^2-3 e^{2   (\varphi _2+\varphi _g)} \chi _g^2\notag\\
   &+6 e^{\varphi _1+4 \varphi _2+\varphi _g} (1+\chi _2^2)+e^{2 (\varphi _1+\varphi _g)} \chi _g^2+3 e^{2 (\varphi _1+\varphi _2+\varphi _g)} (\chi _1-\chi _2)^2 \chi _g^2\notag\\
   &+3 e^{6 \varphi _2+2 \varphi _g} \chi _2^2 (-1+\chi _2 \chi _g)^2+3 e^{2 (\varphi _1+2 \varphi _2+\varphi _g)} (\chi _1-2\chi _1 \chi _2 \chi _g+\chi _2^2 \chi _g)^2\notag\\
   &\left.+e^{2 (\varphi _1+3 \varphi _2+\varphi _g)} [1+\chi _2^3 \chi _g-3 \chi _1 \chi _2 (-1+\chi _2 \chi _g)]^2\right].
\end{align}
This potential admits a trivial critical point at which all scalars vanish. The cosmological constant is given by
\begin{equation}
V_0=-\frac{3}{8}\lambda^2\, .
\end{equation}
At this critical point, we find the scalar masses as in table \ref{table1}. In the table, we also give the corresponding dimensions of the dual operators. The $AdS_4$ radius is given by $L=\frac{2\sqrt{2}}{\lambda}$. Note that we have used different convention for scalar masses from that used in \cite{type_II_orbifold}. The masses given in table \ref{table1} are obtained by multiplying the masses given in \cite{type_II_orbifold} by $3$.
\\
\begin{table}[h]
\centering
\begin{tabular}{|c|c|c|}
  \hline
  Scalar fields & $m^2L^2\phantom{\frac{1}{2}}$ & $\Delta$  \\ \hline
  $\varphi_g$, $\chi_g$ & $-2$ &  $1,2$  \\
  $\varphi_1$, $\varphi_2$ & $4$ &  $4$  \\
   $\chi_1$, $\chi_2$ & $0$ &  $3$  \\
  \hline
\end{tabular}
\caption{Scalar masses at the $N=4$ supersymmetric $AdS_4$ critical
point with $SO(3)\times SO(3)$ symmetry and the
corresponding dimensions of the dual operators}\label{table1}
\end{table}
\indent This $AdS_4$ critical point preserves $N=4$ supersymmetry as can be checked from the $A^{ij}_1$ tensor. It should also be emphasized here that this critical point has $SO(3)\times SO(3)$ symmetry which is the maximal compact subgroup of $ISO(3)\times ISO(3)$ gauge group.
\\
\indent
To set up the BPS equations for finding supersymmetric RG flow solutions, we first give the metric ansatz
\begin{equation}
ds^2=e^{2A(r)}dx^2_{1,2}+dr^2
 \end{equation}
where $dx^2_{1,2}$ is the flat Minkowski metric in three dimensions.
\\
\indent We will use the Majorana representation for gamma matrices with all $\gamma^\mu$ real and $\gamma_5$ purely imaginary. This choice implies that $\epsilon_i$ is a complex conjugate of $\epsilon^i$. All scalar fields will be functions of only the radial coordinate $r$. To solve the BPS conditions coming from setting $\delta\chi^i=0$ and $\delta \lambda^i_a=0$, we need the following projection
\begin{equation}
\gamma_{\hat{r}}\epsilon^i=e^{i\Lambda}\epsilon^i\, .\label{gamma_r_projector}
\end{equation}
From the $\delta\psi_{\mu i}=0$ conditions for $\mu=0,1,2$, we find
\begin{equation}
A'=\pm W,\qquad e^{i\Lambda}=\pm \frac{\mc{W}}{W}
\end{equation}
where $W=|{\mc{W}}|$, and $'$ denotes the $r$-derivative. These equations are obtained by solving real and imaginary parts of $\delta \psi_{\mu i}=0$ separately, see more details in \cite{N3_SU2_SU3,N3_Janus}. The superpotential $\mc{W}$ is defined by
\begin{equation}
\mc{W}=\frac{1}{3}\alpha
\end{equation}
where $\alpha$ is the eigenvalue of $A^{ij}_1$ corresponding to the unbroken supersymmetry. We will choose a definite sign for the $A'$ equation and $e^{i\Lambda}$ such that the $N=4$ critical point identified with
the $N=4$ SCFT in the UV corresponds to $r\rightarrow \infty$.
\\
\indent For all scalars non-vanishing, the $N=4$ supersymmetry is broken to $N=1$ corresponding to the Killing spinor $\epsilon^1$. The superpotential for this unbroken $N=1$ supersymmetry is given by
\begin{align}
\mathcal{W} =& \frac{1}{4 \sqrt{2}}e^{\frac{1}{2} (\varphi _1-3 \varphi _2-\varphi _g)} \left[
e^{\varphi _2} [e^{\varphi _2+\varphi _g} (-e^{\varphi _1+\varphi _2} \lambda -3 \lambda    (i+e^{\varphi _1} \chi _1) (i+e^{\varphi _2} \chi _2))
\right.\notag\\
&\left.-e^{\varphi _1} \lambda  (i+e^{\varphi _2} \chi _2)^3   (i+e^{\varphi _g} \chi _g)+3 \lambda  (i+e^{\varphi _1} \chi _1)   (i+e^{\varphi _2} \chi _2)^2 (i+e^{\varphi _g} \chi _g)]\right]
\end{align}
from which we find
\begin{align}
W =& \frac{1}{8 \sqrt{2}}\lambda e^{\frac{1}{2}(\varphi _1-3 \varphi _2-\varphi _g)} \left[[(-3 e^{\varphi _2} (-e^{\varphi _1}+2   e^{\varphi _2}+e^{2 \varphi _2+\varphi _g}) \chi _2-e^{\varphi _1+3 \varphi _2} \chi _2^3\right.\notag\\
   &+e^{\varphi _g}   (e^{\varphi _1}-3 e^{\varphi _2}) \chi _g+3 e^{2 \varphi _2+\varphi _g} (-e^{\varphi _1}+e^{\varphi   _2}) \chi _2^2 \chi _g+3 e^{\varphi _1+\varphi _2} \chi _1 (-1\notag\\
   &-e^{\varphi _2+\varphi _g}+e^{2 \varphi _2} \chi   _2^2+2 e^{\varphi _2+\varphi _g} \chi _2 \chi _g))^2+[e^{\varphi _1} (-1+3 e^{2 \varphi _2} \chi   _2^2)\notag\\
   &-e^{\varphi _1+\varphi _2+\varphi _g} \chi _2 (-3+e^{2 \varphi _2} \chi _2^2) \chi _g+e^{\varphi _2}   (3+3 e^{\varphi _2+\varphi _g}-e^{\varphi _1+2 \varphi _2+\varphi _g}\notag\\
   &-3 e^{2 \varphi _2} \chi _2^2-6 e^{\varphi   _2+\varphi _g} \chi _2 \chi _g+3 e^{\varphi _1} \chi _1 (-e^{\varphi _2} (2+e^{\varphi _2+\varphi _g}) \chi   _2\notag\\
   &\left.-e^{\varphi _g} \chi _g+e^{2 \varphi _2+\varphi _g} \chi _2^2 \chi _g))]^2]\right]^{\frac{1}{2}}\, .\label{IIB_W}
\end{align}
The scalar potential can be written in term of $W$ as
\begin{equation}
V=-2K^{ij}\frac{\pd W}{\pd \phi^i}\frac{\pd W}{\pd \phi^j}-3W^2,
\end{equation}
and, as usual, the BPS equations from $\delta\chi^i=0$ and $\delta\lambda^i_a=0$ can be written as
\begin{equation}
{\phi^i}'=2K^{ij}\frac{\pd W}{\pd \phi^j}.
\end{equation}
$K^{ij}$ is the inverse of the scalar kinetic metric defined in \eqref{scalar_kin}. The explicit form of these equations is rather complicated and will not be given here. However, they can be found in the appendix.
\\
\indent It is also straightforward to check that these BPS equations solve the second order field equations. Furthermore, there exist a number of interesting subtruncations keeping some subsets of these $SO(3)$ singlets. We will firstly discuss these truncations and consider the full $SO(3)$ singlet sector at the end of this section.

\subsection{RG flows with $N=4$ supersymmetry}
We begin with RG flow solutions preserving $N=4$ supersymmetry to $N=4$ non-conformal field theories in the IR. The analysis of BPS conditions $\delta\psi_{\mu i}=0$, $\delta \chi^i=0$ and $\delta \lambda^i_a=0$ shows that there are two possibilities in order to preserve $N=4$ supersymmetry. The first one is to truncate out $\varphi_{1,2}$ and $\chi_{1,2}$. The second possibility is to keep only the three dilatons $\varphi_{g}$ and $\varphi_{1,2}$ by setting $\chi_{g}=\chi_{1,2}=0$.

\subsubsection{$N=4$ RG flows by relevant deformations}
From table \ref{table1}, we see that $(\varphi_g,\chi_g)$ correspond to relevant deformations by operators of dimensions $1$ or $2$. The BPS equations admit a consistent truncation to these two scalars. By setting $\varphi_{1,2}=\chi_{1,2}=0$, we obtain a set of simple BPS equations
\begin{eqnarray}
\varphi_g'&=&-\frac{\lambda e^{-\frac{\varphi_g}{2}}}{2\sqrt{2}}\frac{(e^{2\varphi_g}+e^{2\varphi_g}\chi_g^2-1)}{\sqrt{(1+e^{\varphi_g})^2+e^{2\varphi_g}\chi_g^2}},\label{N4_eq1}\\
\chi_g'&=&-\frac{\lambda e^{-\frac{\varphi_g}{2}}}{\sqrt{2}}\frac{\chi_g}{\sqrt{(1+e^{\varphi_g})^2+e^{2\varphi_g}\chi_g^2}},\label{N4_eq2}\\
A'&=&\frac{\lambda e^{-\frac{\varphi_g}{2}}}{4\sqrt{2}}\sqrt{(1+e^{\varphi_g})^2+e^{2\varphi_g}\chi_g^2}\, .\label{N4_eq3}
\end{eqnarray}
Since $(\varphi_g,\chi_g)$ are scalars in $SL(2,\mathbb{R})/SO(2)$, they are $SO(6,6)$ singlets and hence $SO(3)\times SO(3)$ invariant. All solutions to these equations then preserve the full $SO(3)\times SO(3)$ symmetry. Moreover, equations $\delta\lambda^i_a=0$ are identically satisfied, and it can be checked that $N=4$ supersymmetry is unbroken since equations $\delta\chi^i=0$ and $\delta\psi_{\mu i}=0$ hold for all $\epsilon^i$ satisfying the $\gamma_{\hat{r}}$ projector \eqref{gamma_r_projector}. We should clarify here the convention on the number of supersymmetry. In four dimensions, the $\gamma_{\hat{r}}$ projector reduce the number of supercharges from $16$ to $8$. The latter corresponds to $N=4$ supersymmetry in three dimensions. On the other hand, the $AdS_4$ vacuum preserves all $16$ supercharges corresponding to $N=4$ superconformal symmetry in three dimensions containing $8+8=16$ supercharges.
\\
\indent We begin with an even simpler solution with $\chi_g=0$ which, from the above equations, is clearly a consistent truncation. In this case, we end up with the BPS equations
\begin{eqnarray}
\varphi_g'&=&-\frac{\lambda}{2\sqrt{2}}e^{-\frac{\varphi_g}{2}}(e^{\varphi_g}-1),\\
A'&=&\frac{\lambda}{4\sqrt{2}}e^{-\frac{\varphi_g}{2}}(1+e^{\varphi_g}).
\end{eqnarray}
The solution to these equations is easily found to be
\begin{eqnarray}
\varphi_g &=&\ln \left[e^{\frac{r\lambda}{2\sqrt{2}}+C}-1\right]-\ln \left[e^{\frac{r\lambda}{2\sqrt{2}}+C}+1\right],\\
A&=&\ln \left[e^{\frac{r\lambda}{2\sqrt{2}}+C}-1\right]-\frac{r\lambda}{2\sqrt{2}}
\end{eqnarray}
with $C$ being an integration constant. The additive integration constant for $A$ has been neglected since it can be absorbed by scaling $dx^2_{1,2}$ coordinates. In addition, the constant $C$ can also be removed by shifting the $r$ coordinate.
\\
\indent At large $r$, we find, as expected for dual operators of dimensions $\Delta=1,2$,
\begin{equation}
\varphi_g\sim e^{-\frac{\lambda r}{2\sqrt{2}}}\sim e^{-\frac{r}{L}}\, .
\end{equation}
The solution is singular as $r\rightarrow -\frac{2\sqrt{2}C}{\lambda}$ since $\varphi_g\rightarrow -\infty$. Near this singularity, we find
\begin{equation}
\varphi_g\sim A\sim \ln \left[r+\frac{2\sqrt{2}C}{\lambda}\right].
\end{equation}
The scalar potential is bounded above with $V\rightarrow -\infty$. Therefore, the singularity of this solution is physical by the criterion given in \cite{Gubser_singularity}. The solution then describes an RG flow from the dual $N=4$ SCFT to an $N=4$ non-conformal field theory with unbroken $SO(3)\times SO(3)$ symmetry. The metric in the IR is given by
\begin{equation}
ds^2=(\lambda r+2\sqrt{2}C)^2dx^2_{1,2}+dr^2
\end{equation}
where we have absorbed some constants to $dx^2_{1,2}$ coordinates.
\\
\indent We then consider possible flows solution with $\chi_g\neq 0$. By introducing a new variable $\rho$ via
\begin{equation}
\frac{d\rho}{dr} =\frac{\chi_g}{\sqrt{1-C\chi_g}+\sqrt{1-\chi_g(2C+\chi_g)}}
\end{equation}
we find the following solution to equations \eqref{N4_eq1},\eqref{N4_eq2} and \eqref{N4_eq3}
\begin{eqnarray}
\varphi_g&=&-\frac{1}{2}\ln [1-2C\chi_g-\chi_g^2],\label{N4_2sol1}\\
A&=&-\ln \chi_g+\frac{1}{4}\ln [1-2C\chi_g-\chi_g^2]\nonumber \\
& &+\frac{1}{2}\ln \left[1-C\chi_g+\sqrt{1-2C\chi_g-\chi_g^2}\right],\label{N4_2sol2}\\
\rho\lambda [1-\chi_g(2C+\chi_g)]^\frac{3}{4}&=&4(2)^{\frac{1}{4}}(C+\chi_g-\sqrt{1+C^2})\left[\frac{1+C^2+\sqrt{1+C^2}(C+\chi_g)}{1+C^2}\right]^{\frac{3}{4}}
\times\nonumber \\& &_2F_1\left(\frac{1}{4},\frac{3}{4},\frac{5}{4},\frac{\chi_g+\sqrt{1+C^2}-C}{2\sqrt{1+C^2}}\right)\label{N4_2sol3}
\end{eqnarray}
where $_2F_1$ is the hypergeometric function.
\\
\indent The solution interpolates between the $N=4$ $AdS_4$ vacuum as $r\rightarrow \infty$ and a singular geometry at a finite value of $r$. There are two possibilities for the IR singularities. The first one is given by
\begin{eqnarray}
\chi_g &\sim &\chi_0,\qquad \varphi_g\sim -2\ln\left[ \frac{\sqrt{2}r\lambda(1+\chi_0^2)-4\chi_0C}{8\chi_0}\right],\nonumber \\
A &\sim & \frac{\chi_0}{\sqrt{1+\chi_0}}\ln[\sqrt{2}r\lambda(1+\chi_0^2)-4\chi_0C]
\end{eqnarray}
where $\chi_0$ is a constant. In this case, we have $\varphi_g\rightarrow \infty$ and $\chi_g\rightarrow \textrm{constant}$ as $\sqrt{2}\lambda r(1+\chi_0^2)\rightarrow 4\chi_0C$. It should be noted here that the constant $C$ in these equations is not the same as in the full solution given in \eqref{N4_2sol1}, \eqref{N4_2sol2} and \eqref{N4_2sol3}.
\\
\indent Another possibility is given by
\begin{eqnarray}
\varphi_g&\sim &2\ln(\sqrt{2}\lambda r+4C),\qquad \chi_g\sim \frac{\tilde{C}}{4C+\sqrt{2}\lambda r},\nonumber \\
A&\sim &\ln (\sqrt{2}\lambda r+4C)\, .
\end{eqnarray}
In this case, as $\sqrt{2}r\lambda\rightarrow -4C$, we have $\varphi_g\rightarrow -\infty$ and $\chi_g\rightarrow \pm \infty$ depending on the sign of the constant $\tilde{C}$. Both of these singularities lead to $V\rightarrow -\infty$, so they are physical.

\subsubsection{$N=4$ RG flows by relevant and irrelevant deformations}
We now consider RG flows with $N=4$ supersymmetry with $\chi_g=\chi_{1,2}=0$. Recall that $\varphi_{1}$ and $\varphi_2$ are $SO(3)\times SO(3)$ singlets, we still have solutions with $SO(3)\times SO(3)$ unbroken along the flows. It should also be noted that the truncation $\chi_{1,2}=0$ is consistent only for $\chi_g=0$. This implies that $N=4$ supersymmetry does not allow turning on the operators dual to $\chi_g$ and $\varphi_{1,2}$ simultaneously. It would be interesting to see the implication of this in the dual $N=4$ SCFT.
\\
\indent In this case, the BPS equations reduce to
\begin{eqnarray}
\varphi_g'&=&\frac{\lambda}{4\sqrt{2}}e^{\frac{1}{2}(\varphi_1-3\varphi_2-\varphi_g)}
\left(3e^{\varphi_2}-e^{\varphi_1}-3e^{2\varphi_2+\varphi_g}+e^{\varphi_1+3\varphi_2+\varphi_g}\right),\\
\varphi_1'&=&\frac{\lambda}{4\sqrt{2}}\left(e^{\varphi_1}-e^{\varphi_2}-e^{2\varphi_2+\varphi_g}
+e^{\varphi_1+3\varphi_2+\varphi_g}\right),\\
\varphi_2'&=&\frac{\lambda}{4\sqrt{2}}\left(e^{\varphi_2}-e^{\varphi_1}-e^{2\varphi_2+\varphi_g}
+e^{\varphi_1+3\varphi_2+\varphi_g}\right),\\
A'&=&-\frac{\lambda}{8\sqrt{2}}e^{\frac{1}{2}(\varphi_1-3\varphi_2-\varphi_g)}\left(e^{\varphi_1}-3e^{\varphi_2}
-3e^{2\varphi_2+\varphi_g}+e^{\varphi_1+3\varphi_2+\varphi_g}\right).
\end{eqnarray}
These equations can be analytically solved by introducing new variables
\begin{equation}
\tilde{\varphi}_1=\varphi_1-\varphi_2,\qquad \tilde{\varphi}_2=\varphi_1+\varphi_2
\end{equation}
in terms of which the BPS equations become
\begin{eqnarray}
\tilde{\varphi}_1'&=&\frac{\lambda}{2\sqrt{2}}e^{\frac{1}{2}(\tilde{\varphi}_1+\varphi_g)}
(e^{\tilde{\varphi}_1}-1),\label{N42_eq1}\\
\tilde{\varphi}_2'&=&\frac{\lambda}{2\sqrt{2}}e^{\frac{1}{2}(\tilde{\varphi}_2-\varphi_g)}
(e^{\tilde{\varphi}_2}-1),\label{N42_eq2}\\
\varphi_g'&=&\frac{\lambda}{4\sqrt{2}}e^{-\frac{\varphi_g}{2}}\left(3e^{\frac{\tilde{\varphi}_2}{2}}
-e^{\frac{3}{2}\tilde{\varphi}_2}-3e^{\frac{1}{2}\tilde{\varphi}_1+\varphi_g}
+e^{\frac{3}{2}\tilde{\varphi}_1+\varphi_g}\right),\label{N42_eq3}\\
A'&=&-\frac{\lambda}{8\sqrt{2}}e^{-\frac{\varphi_g}{2}}\left(e^{\frac{3}{2}\tilde{\varphi}_1+\varphi_g}
+e^{\frac{3}{2}\tilde{\varphi}_2}-3e^{\frac{\tilde{\varphi}_2}{2}}-3e^{\frac{1}{2}\tilde{\varphi}_1
+\varphi_g}\right).\label{N42_eq4}
\end{eqnarray}
By combining all of these equations, we find that
\begin{eqnarray}
\frac{dA}{d\tilde{\varphi}_1}-\frac{1}{2}\frac{d\varphi_g}{d\tilde{\varphi}_1}&=&
\frac{3-e^{\tilde{\varphi}_1}}{2(e^{\tilde{\varphi}_1}-1)},\\
\frac{dA}{d\tilde{\varphi}_2}+\frac{1}{2}\frac{d\varphi_g}{d\tilde{\varphi}_2}&=&
\frac{3-e^{\tilde{\varphi}_2}}{2(e^{\tilde{\varphi}_2}-1)}
\end{eqnarray}
which can be solved by the following solution
\begin{eqnarray}
\varphi_g&=&\frac{3}{2}(\tilde{\varphi}_1-\tilde{\varphi}_2)-\ln(1-e^{\tilde{\varphi}_1})
+\ln(1-e^{\tilde{\varphi}_2}),\\
A&=&\frac{\varphi_g}{2}-\frac{3}{2}\tilde{\varphi}_1+\ln(1-e^{\tilde{\varphi}_1}).
\end{eqnarray}
In this solution, we have fixed the integration constant for $\varphi_g$ to zero since at the $AdS_4$ critical point $\varphi_g=\tilde{\varphi}_1=\tilde{\varphi}_2=0$. The integration constant for $A$ is irrelevant.
\\
\indent
Combining equations \eqref{N42_eq1} and \eqref{N42_eq2}, we obtain after substituting for $\varphi_g$
\begin{equation}
\frac{d\tilde{\varphi}_1}{d\tilde{\varphi}_2}=e^{2(\tilde{\varphi}_1-\tilde{\varphi}_2)}
\end{equation}
whose solution is given by
\begin{equation}
\tilde{\varphi}_1=-\frac{1}{2}\ln (e^{-2\tilde{\varphi}_2}-C_1).
\end{equation}
Near the $AdS_4$ critical point, we have $\tilde{\varphi}_1\sim \tilde{\varphi}_2\sim 0$ which requires that $C_1=0$. This choice leads to $\tilde{\varphi}_2=\tilde{\varphi}_1$ which implies $\varphi_2=0$ and $\varphi_g=0$. We see that the flow does not involve $\varphi_g$ and is driven purely by an irrelevant operator of dimension $4$ dual to $\varphi_1$. In this case, the $N=4$ SCFT dual to the $AdS_4$ vacuum is expected to appear in the IR. Note also that equation \eqref{N42_eq3} is consistent for $\varphi_g=0$ if and only if $\tilde{\varphi}_2=\tilde{\varphi}_1$ as being the case here.
\\
\indent Finally, we can solve equation \eqref{N42_eq1} for $\tilde{\varphi}_1(r)$
\begin{equation}
\frac{\lambda r}{2\sqrt{2}}=2e^{-\frac{\tilde{\varphi}_1}{2}}+\ln (1-e^{-\frac{\tilde{\varphi}_1}{2}})-
\ln (1+e^{-\frac{\tilde{\varphi}_1}{2}})+C\, .
\end{equation}
The solution is singular as $r\rightarrow \frac{2\sqrt{2}C}{\lambda}$. Near this singularity, the solution becomes
\begin{eqnarray}
\tilde{\varphi}_1&\sim &\tilde{\varphi}_2\sim -\frac{2}{3}\ln \frac{3}{2}\left[C-\frac{\lambda r}{2\sqrt{2}}\right],\nonumber \\
A&\sim &-\frac{1}{2}\tilde{\varphi}_1\sim \frac{1}{3}\ln \frac{3}{2}\left[C-\frac{\lambda r}{2\sqrt{2}}\right].
\end{eqnarray}
This singularity leads to $V\rightarrow \infty$, so the solution is unphysical.
\\
\indent We end the discussion of this truncation by giving some comments on possible subtruncations. From equations \eqref{N42_eq1} and \eqref{N42_eq2}, we easily see that setting $\tilde{\varphi}_1=0$ or $\tilde{\varphi}_2=0$ is a consistent truncation. This is equivalent to setting $\varphi_2=\pm \varphi_1$. In this case, the solution is found to be
\begin{eqnarray}
\varphi_g&=&\pm \ln\left[\frac{e^{-\varphi_1}-C_1e^{3\varphi_1}}{2-2e^{2\varphi_1}}\right],\nonumber \\
A&=&-\frac{7}{2}\varphi_1+\frac{1}{2}\ln (1-e^{2\varphi_1})+\frac{1}{2}\ln (1-C_1e^{4\varphi_1}),\nonumber \\
\frac{\lambda \rho}{4\sqrt{2}}&=&e^{-\varphi_1}+\frac{1}{2}\ln (1-e^{-\varphi_1})-\frac{1}{2}\ln(1+e^{-\varphi_1})+C
\end{eqnarray}
where the new radial coordinate $\rho$ is defined by $d\rho=e^{-\frac{\varphi_g}{2}}dr$.
\\
\indent We see that in this case $\varphi_g$ is non-trivial along the flow. In order to make the solution approach the $AdS_4$ critical point with $\varphi_g\sim \varphi_1\sim 0$, we need to choose $C_1=1$. This gives
\begin{equation}
\varphi_g=\pm \ln\cosh \varphi_1\, .
\end{equation}
The solution is singular for $\rho\rightarrow \frac{4\sqrt{2}C}{3\lambda}$ with $\varphi_1\rightarrow \infty$. In this limit, we find
\begin{equation}
\varphi_g\sim \pm \varphi_1\qquad \varphi_1\sim -\ln\left[C-\frac{3\lambda \rho}{4\sqrt{2}}\right],\qquad
A\sim \frac{1}{2}\ln\left[C-\frac{3\lambda \rho}{4\sqrt{2}}\right].
\end{equation}
Both of these singularities lead to $V\rightarrow \infty$, so they are also unphysical.

\subsection{RG flows with $N=1$ supersymmetry}
We now consider a class of RG flow solutions preserving $N=1$ supersymmetry and breaking the $SO(3)\times SO(3)$ to its diagonal subgroup. This is achieved by turning on the marginal deformations corresponding to $\chi_1$ and $\chi_2$ to the solutions. As in the $N=4$ case, there is a consistent subtruncation to only irrelevant and marginal deformations with only $\varphi_1$ and $\chi_1$ non-vanishing. We will consider this case first and then look for the most general solutions with all six $SO(3)$ singlet scalars non-vanishing. It should be noted that the truncation with only $\varphi_2$ and $\chi_2$ non-vanishing is not consistent. This is also an interesting feature to look for in the dual field theory.

\subsubsection{$N=1$ RG flows by marginal and irrelevant deformations}
\begin{figure}
         \centering
         \begin{subfigure}[b]{0.3\textwidth}
                 \includegraphics[width=\textwidth]{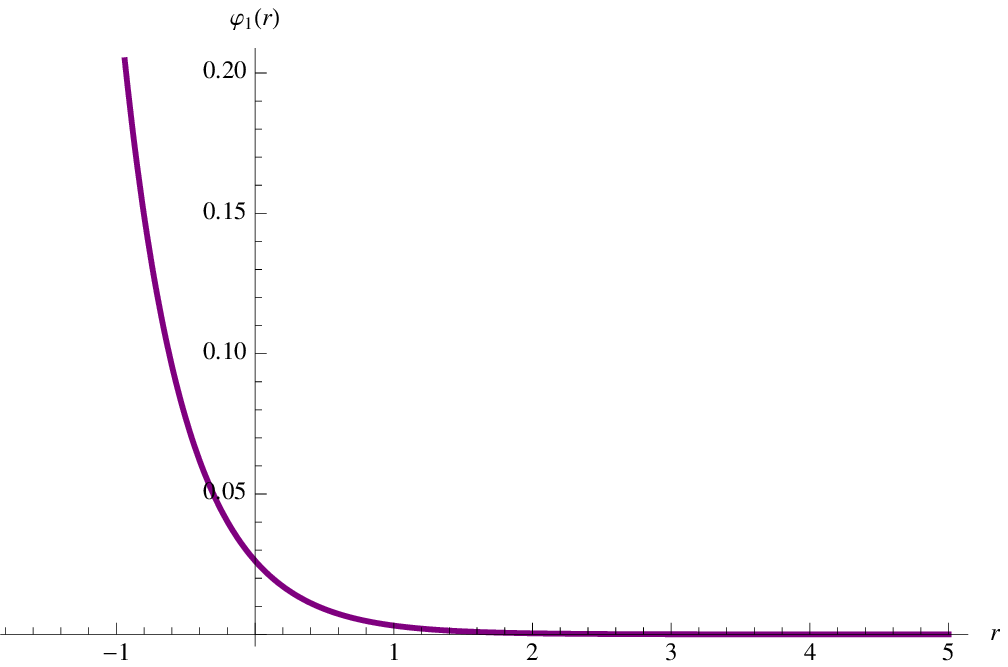}
                 \caption{Solution for $\varphi_1$}
         \end{subfigure}
\begin{subfigure}[b]{0.3\textwidth}
                 \includegraphics[width=\textwidth]{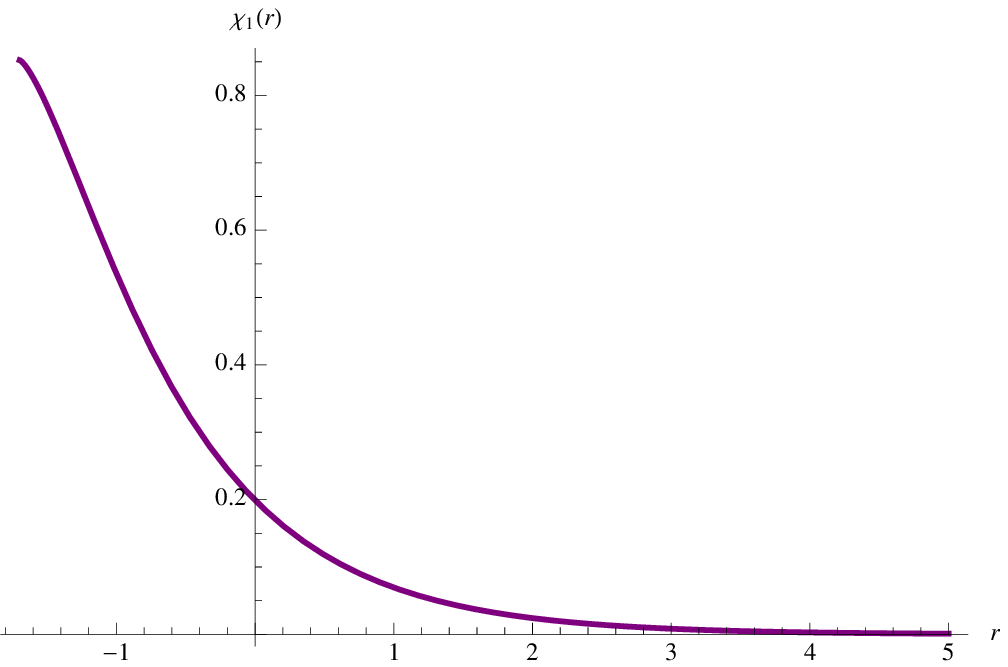}
                 \caption{Solution for $\chi_1$}
         \end{subfigure}%
         ~ %add desired spacing between images, e. g. ~, \quad, \qquad, \hfill etc.
         \begin{subfigure}[b]{0.3\textwidth}
                 \includegraphics[width=\textwidth]{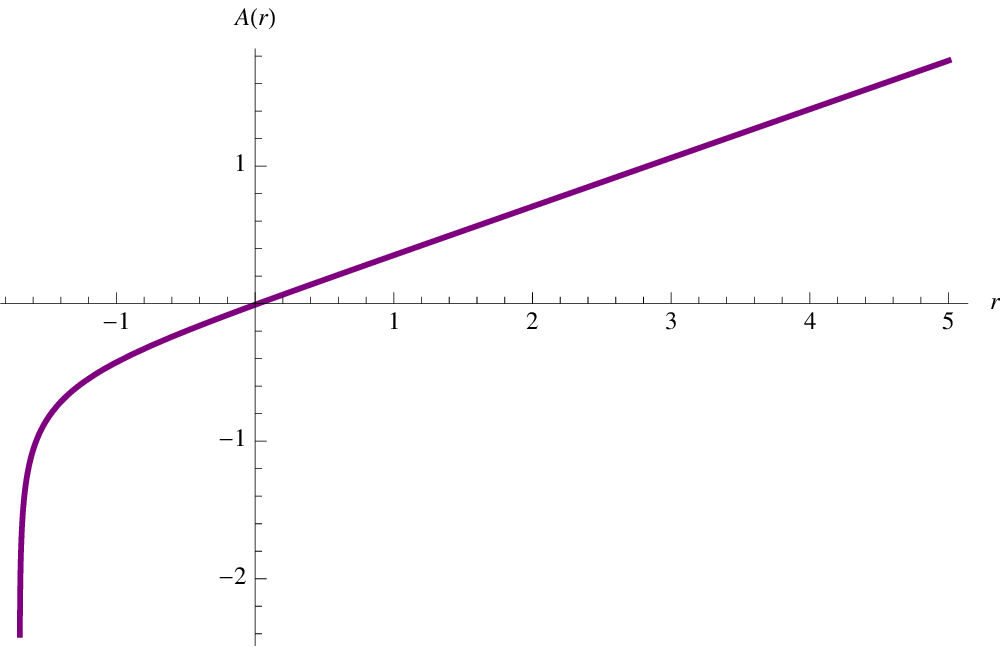}
                 \caption{Solution for $A$}
         \end{subfigure}
         \caption{An $N=1$ RG flow with marginal and irrelevant deformations from type IIB compactification}\label{fig1}
 \end{figure}

By setting $\varphi_g=\chi_g=\varphi_{2}=\chi_2=0$ in the BPS equations, we obtain
\begin{eqnarray}
\varphi_1'&=&-\frac{\lambda}{2\sqrt{2}}e^{\frac{\varphi_1}{2}}\frac{(3-4e^{\varphi_1}+e^{2\varphi_1}+9\chi_1^2e^{2\varphi_1})}
{\sqrt{(e^{\varphi_1}-3)^2+9\chi_1^2e^{2\varphi_1}}},\\
\chi_1'&=&-\frac{3\lambda}{2\sqrt{2}}\frac{\chi_1e^{\frac{\varphi_1}{2}}}{\sqrt{(e^{\varphi_1}-3)^2+9\chi_1^2e^{2\varphi_1}}},\\
A'&=&\frac{\lambda}{4\sqrt{2}}e^{\frac{\varphi_1}{2}}\sqrt{(e^{\varphi_1}-3)^2+9\chi_1^2e^{2\varphi_1}}\, .
\end{eqnarray}
We are not able to analytically solve these equations in full generality, so we will look for numerical solutions in this case.
\\
\indent Note that further truncation to only $\varphi_1$ gives rise to the following BPS equations
\begin{equation}
\varphi_1'=\frac{\lambda}{2\sqrt{2}}e^\frac{\varphi_1}{2}(e^{\varphi_1}-1)\qquad \textrm{and}\qquad
A'=-\frac{\lambda}{4\sqrt{2}}e^{\frac{\varphi_1}{2}}(e^{\varphi_1}-3)
\end{equation}
with the solution
\begin{eqnarray}
A&=&-\frac{3}{2}\varphi_1+\ln (1-e^{\varphi_1}),\nonumber \\
\frac{\lambda r}{2\sqrt{2}}&=&2e^{-\frac{\varphi_1}{2}}+\ln(1-e^{-\frac{\varphi_1}{2}})-\ln(1+e^{-\frac{\varphi_1}{2}}).
\end{eqnarray}
This is nothing but the same solution as in the previous section for $\tilde{\varphi}_2=\tilde{\varphi}_1$.
Therefore, we will not further discuss this solution.
\\
\indent For non-vanishing $\chi_1$, we need to find the solutions numerically. An example of these solutions is given in figure \ref{fig1}.
The asymptotic behavior of this solution can be determined from the BPS equations at large $\varphi_1$ as follow
\begin{eqnarray}
\chi_1&\sim & \chi_0,\qquad \varphi_1\sim -\frac{2}{3}\ln (r\lambda \sqrt{2+18\chi_0^2}-4C_1),\nonumber \\
A&\sim &\frac{1}{3}\ln(r\lambda \sqrt{2+18\chi_0^2}-4C_1)
\end{eqnarray}
where $\chi_0$ is a constant. This singularity leads to $V\rightarrow \infty$ implying that it is unphysical. We have in addition checked this by a numerical analysis which consistently shows a diverging scalar potential near the singularity.

\subsubsection{$N=1$ RG flows by relevant, marginal and irrelevant deformations}
\begin{figure}
         \centering
         \begin{subfigure}[b]{0.3\textwidth}
                 \includegraphics[width=\textwidth]{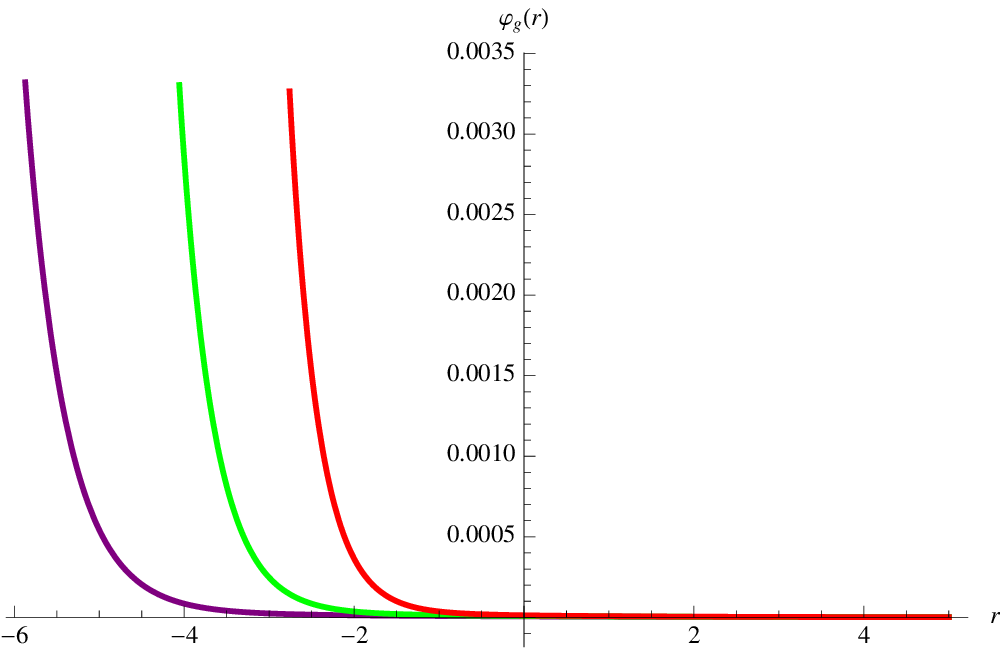}
                 \caption{Solution for $\varphi_g$}
         \end{subfigure}%
         ~ %add desired spacing between images, e. g. ~, \quad, \qquad, \hfill etc.
         \begin{subfigure}[b]{0.3\textwidth}
                 \includegraphics[width=\textwidth]{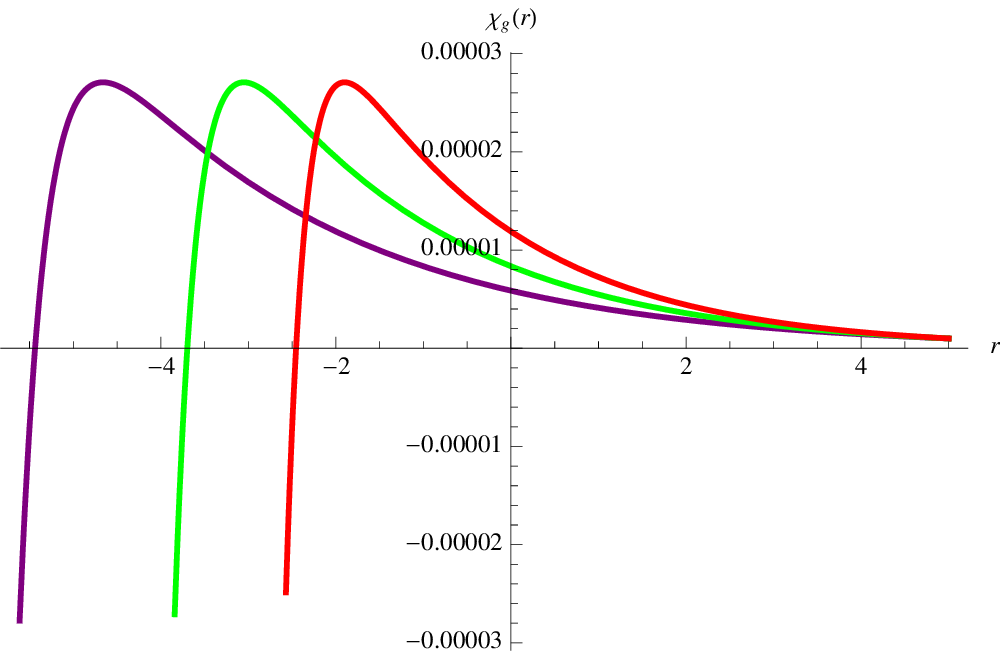}
                 \caption{Solution for $\chi_g$}
         \end{subfigure}
         \begin{subfigure}[b]{0.3\textwidth}
                 \includegraphics[width=\textwidth]{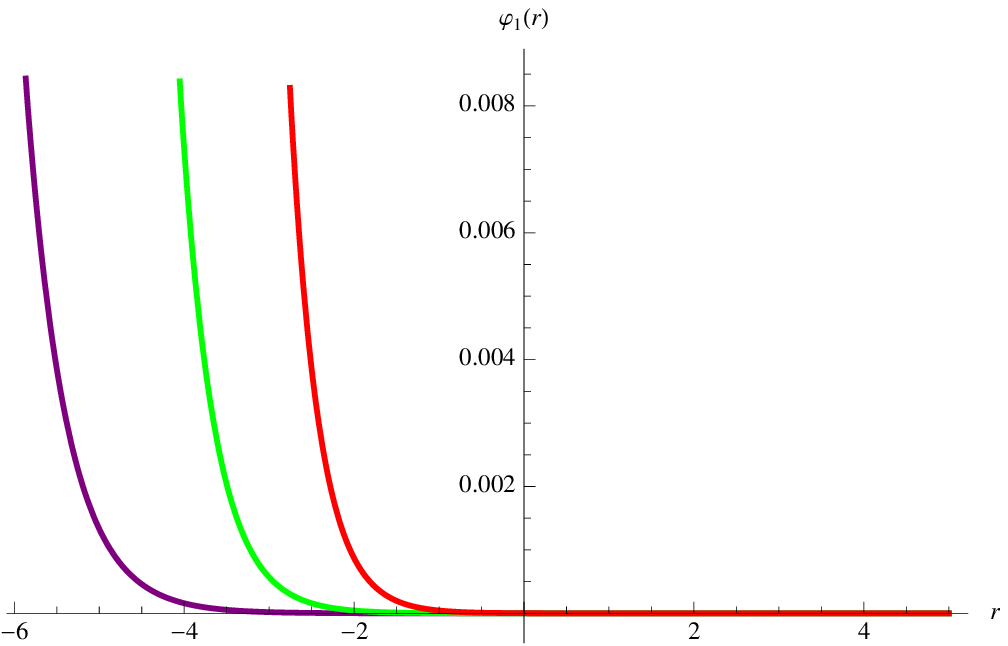}
                 \caption{Solution for $\varphi_1$}
         \end{subfigure}\\
\begin{subfigure}[b]{0.3\textwidth}
                 \includegraphics[width=\textwidth]{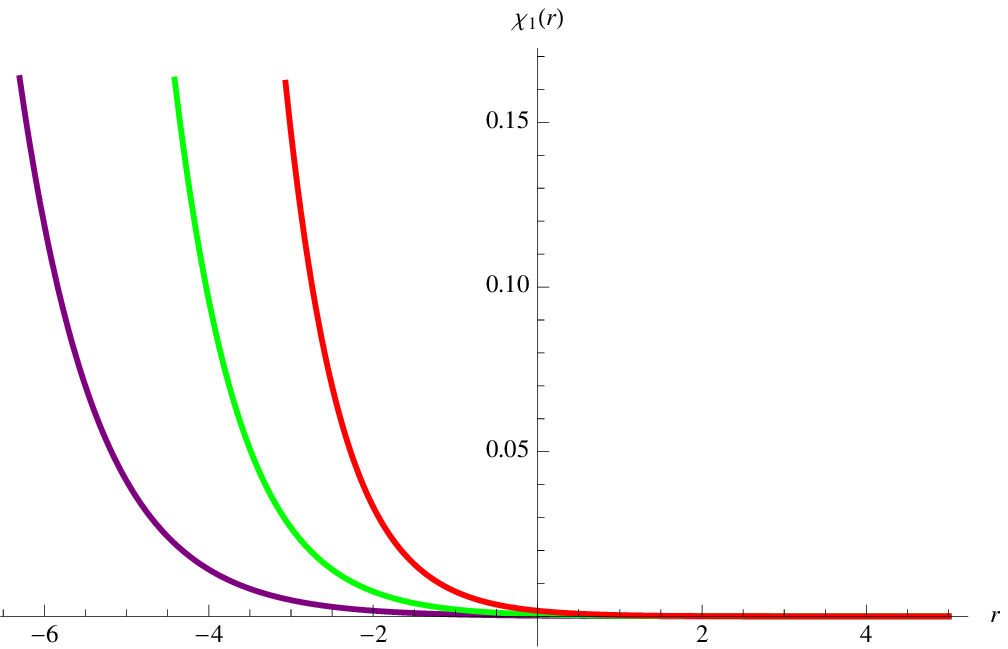}
                 \caption{Solution for $\chi_1$}
         \end{subfigure}%
         ~ %add desired spacing between images, e. g. ~, \quad, \qquad, \hfill etc.
         \begin{subfigure}[b]{0.3\textwidth}
                 \includegraphics[width=\textwidth]{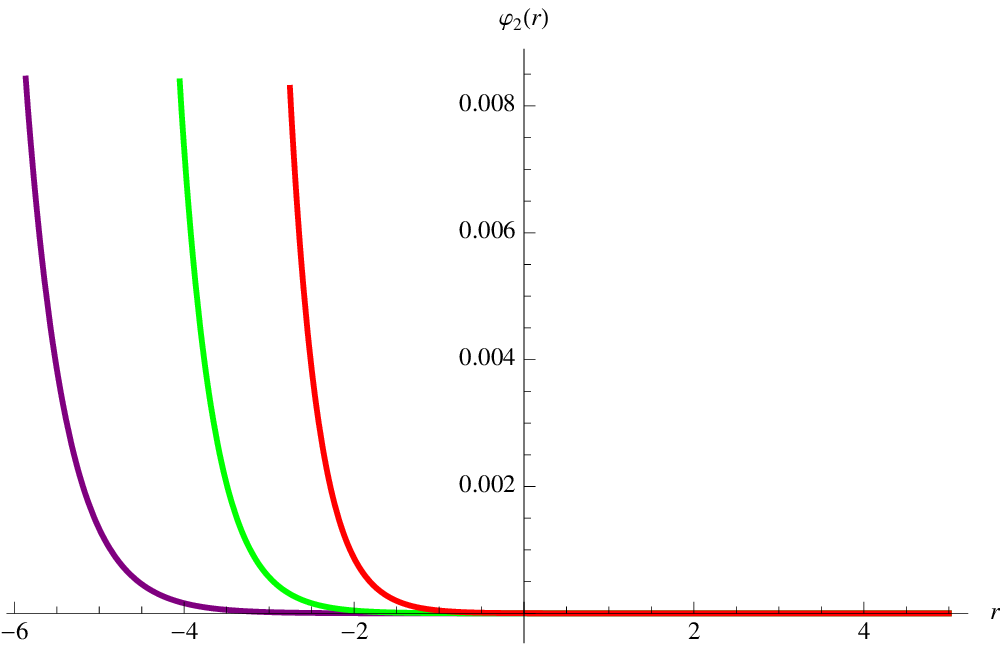}
                 \caption{Solution for $\varphi_2$}
         \end{subfigure}
         \begin{subfigure}[b]{0.3\textwidth}
                 \includegraphics[width=\textwidth]{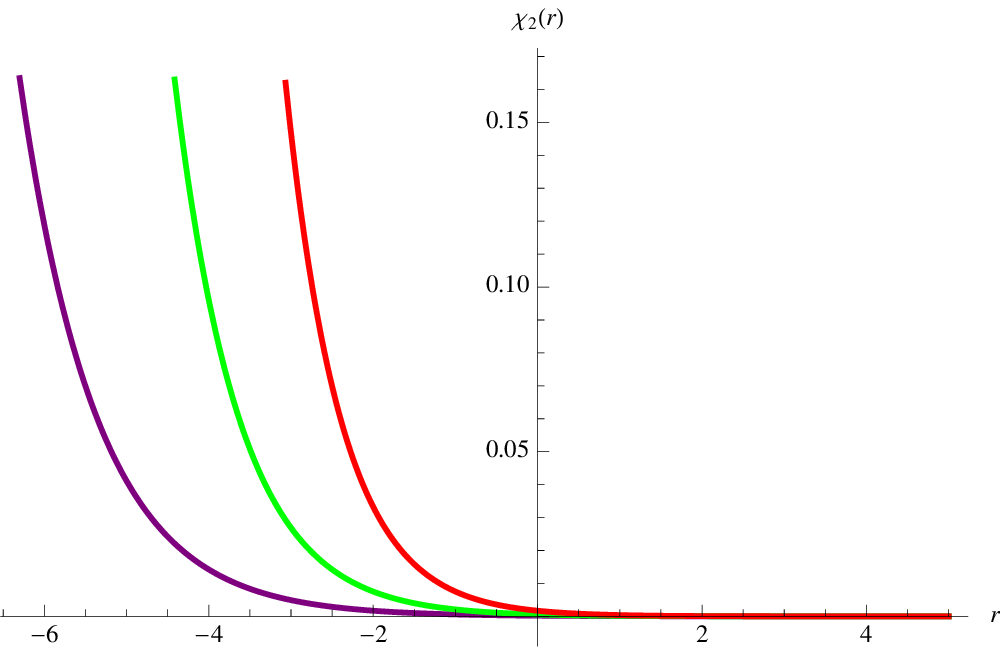}
                 \caption{Solution for $\chi_2$}
         \end{subfigure}\\
         \begin{subfigure}[b]{0.3\textwidth}
                 \includegraphics[width=\textwidth]{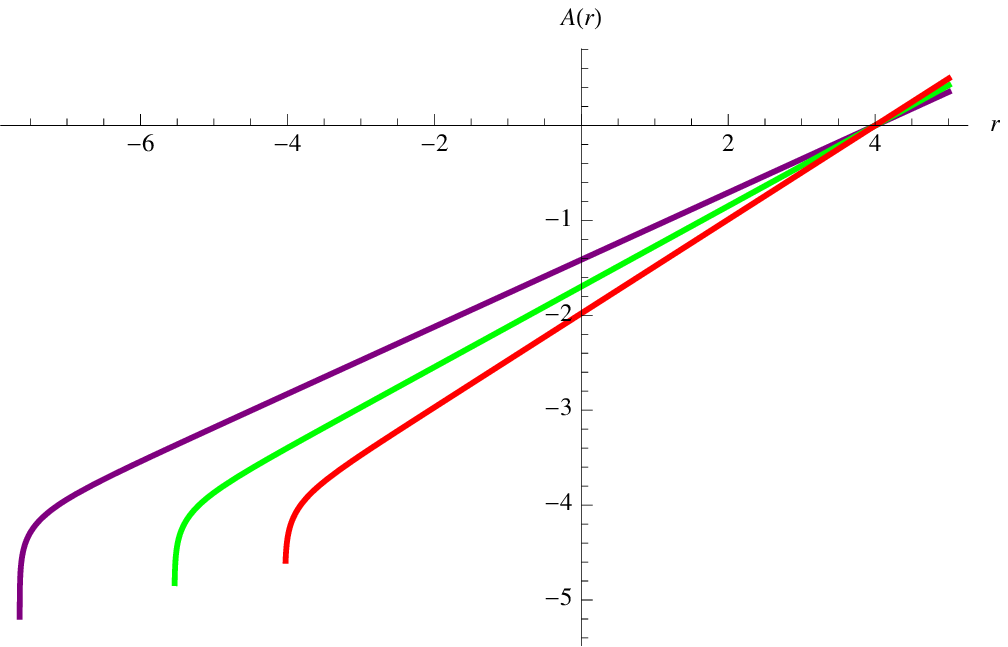}
                 \caption{Solution for $A$}
         \end{subfigure}
         \caption{$N=1$ RG flow solutions from type IIB compactification with all $SO(3)$ singlet scalars and $\lambda=1$ (purple), $\lambda=1.2$ (green) and $\lambda=1.4$ (red)}\label{fig2}
 \end{figure}

After consider various consistent truncations, we end this section by considering $N=1$ RG flow solutions with all six $SO(3)$ singlet scalars turned on. The resulting RG flows will be driven by all types of possible deformations namely marginal, irrelevant and relevant. In this case, we need to use a numerical analysis due to the complexity of the full set of BPS equations given in the appendix. Similar to the analysis of \cite{Yi_4D_flow}, there could be many possible IR singularities due to the competition between various deformations both by operators and vacuum expectation values (vev) present in the UV $N=4$ SCFT. Some examples of these solutions are given in figure \ref{fig2}. In the figure, we have given solutions for three different values of the flux parameter $\lambda$ for comparison.\\
\indent From figure \ref{fig2}, we see a singularity in the IR end of the flow while near $r\rightarrow \infty$ the flow approaches the UV $N=4$ $AdS_4$. The numerical analysis shows that the singularity is of a bad type according to the criterion of \cite{Gubser_singularity} since it leads to $V\rightarrow \infty$.

%%%%%%%%%%%%%%%%%%%%%%%%%%%%%%%%%%%%%%%%%%%%%%%%%%%%%%%%%%%%%%%%%%%%%%%%%%%%%%%%%%%%%%%%%%%%%%%%%%%%%%%%%%%%%%%%%%%%%%%%%%%%%%%%%%%%%%%%%
\section{Supersymmetric Janus solutions}\label{IIB_Janus}
In this section, we look at another type of solutions with an $AdS_3$-sliced domain wall ansatz, obtained by replacing the flat metric $dx^2_{1,2}$ by an $AdS_3$ metric of radius $\ell$,
\begin{equation}
ds^2=e^{2A(r)}\left(e^{\frac{2\xi}{\ell}}dx^2_{1,1}+d\xi^2\right)+dr^2\,
.
\end{equation}
The solution, called Janus solution, describes a conformal interface of co-dimension one within the SCFT dual to the $AdS_4$ critical point. This solution breaks the three-dimensional conformal symmetry $SO(2,3)$ to that on the $(1+1)$-dimensional interface $SO(2,2)$.
\\
\indent
In this case, the resulting BPS equations will get modified compared to the RG flow case. First of all, the analysis of $\delta \psi^i_\mu=0$ equations requires an additional $\gamma_{\hat{\xi}}$ projection
\begin{equation}
\gamma_{\hat{\xi}}\epsilon_i=i\kappa e^{i\Lambda}\epsilon^i
\end{equation}
while the $\gamma_{\hat{r}}$ projector in $\delta\chi^i=0$ and $\delta\lambda^i_a=0$ equations is still given by equation \eqref{gamma_r_projector} but with the phase $e^{i\Lambda}$ modified to
\begin{equation}
e^{i\Lambda}=\frac{\mc{W}}{A'+\frac{i\kappa}{\ell}e^{-A}}\, .\label{Janus_phase}
\end{equation}
Furthermore, the integrability conditions for $\delta
\psi^i_{\hat{0},\hat{1}}=0$ equations lead to
\begin{equation}
A'^2+\frac{1}{\ell^2}e^{-2A}=W^2\, .\label{Janus_gravitino}
\end{equation}
As expected, these two equations reduce to $A'=\pm W$ and $e^{i\Lambda}=\frac{\mc{W}}{A'}=\pm\frac{\mc{W}}{W}$ in the limit $\ell\rightarrow \infty$.
\\
\indent
The constant $\kappa$, with $\kappa^2=1$, imposes the chirality condition on the Killing spinors corresponding to the unbroken supersymmetry on the $(1+1)$-dimensional interface. The detailed analysis of these equations can be found for example in \cite{N3_Janus}. Unlike the RG flow case, the Killing spinors depend on both $r$ and $\xi$ coordinates, see more detail in \cite{warner_Janus}.
\\
\indent We have seen that the analysis of RG flow solutions with all six $SO(3)$ singlet scalars turned on involves a very complicated set of BPS equations. Since the BPS equations for supersymmetric Janus solutions are usually more complicated than those of the RG flows, we will not perform the full analysis with all $SO(3)$ singlet scalars but truncate the BPS equations to two consistent truncations, with $(\varphi_g,\chi_g)$ and $(\varphi_1,\chi_1)$ non-vanishing. As in other cases studied in \cite{tri-sasakian-flow,warner_Janus,N3_Janus}, truncations to only dilatons or scalars without the axions or pseudoscalars are not consistent with the Janus BPS equations, or equivalently Janus solutions require non-trivial pseudoscalars.

\subsection{$N=4$ Janus solution}
\begin{figure}
         \centering
         \begin{subfigure}[b]{0.3\textwidth}
                 \includegraphics[width=\textwidth]{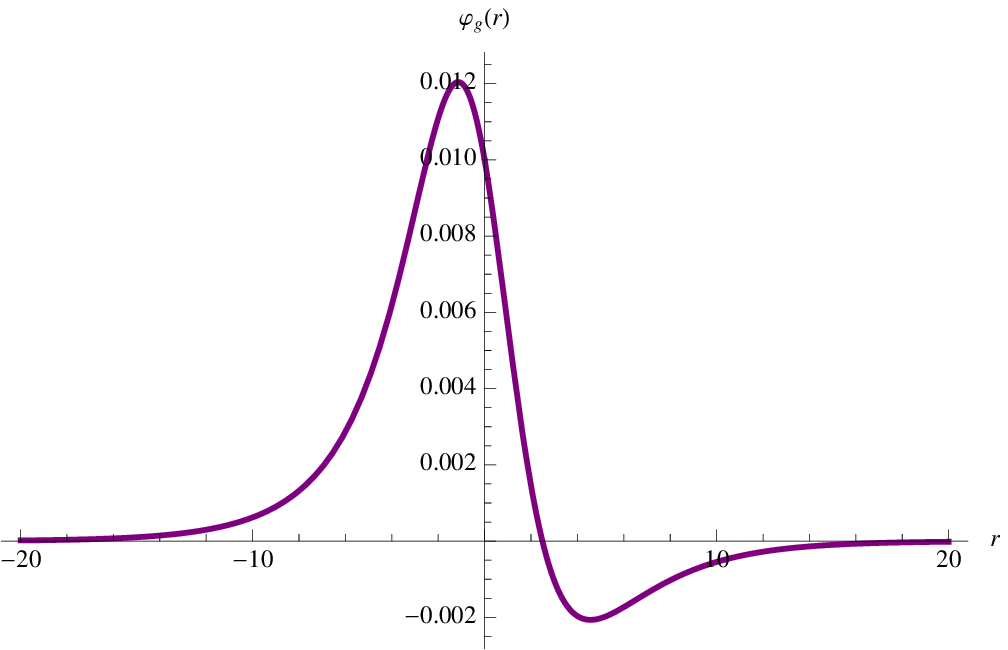}
                 \caption{Solution for $\varphi_g$}
         \end{subfigure}
\begin{subfigure}[b]{0.3\textwidth}
                 \includegraphics[width=\textwidth]{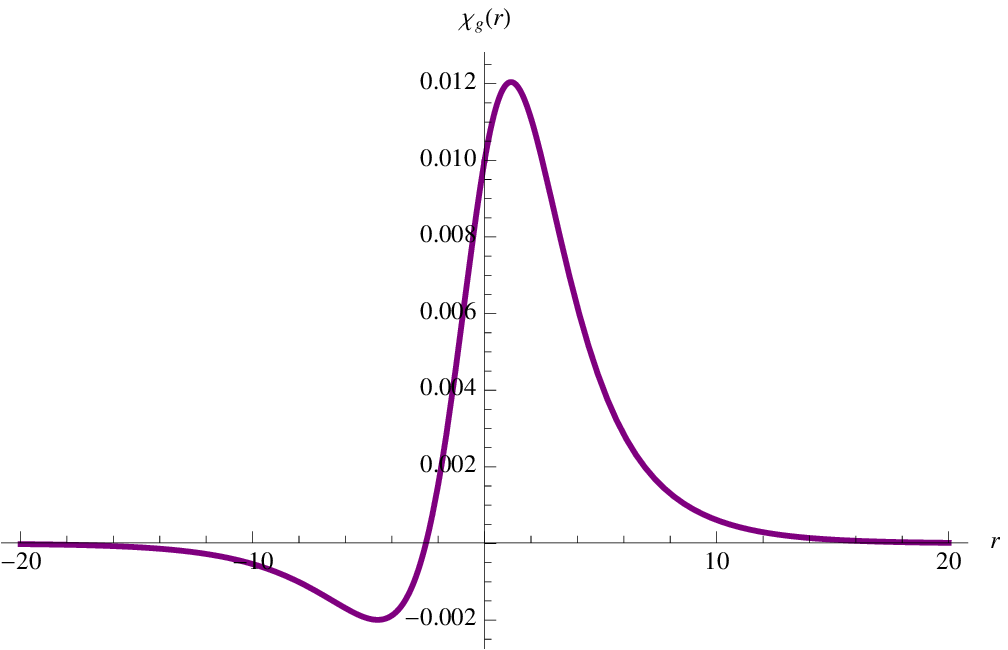}
                 \caption{Solution for $\chi_g$}
         \end{subfigure}%
         ~ %add desired spacing between images, e. g. ~, \quad, \qquad, \hfill etc.
         \begin{subfigure}[b]{0.3\textwidth}
                 \includegraphics[width=\textwidth]{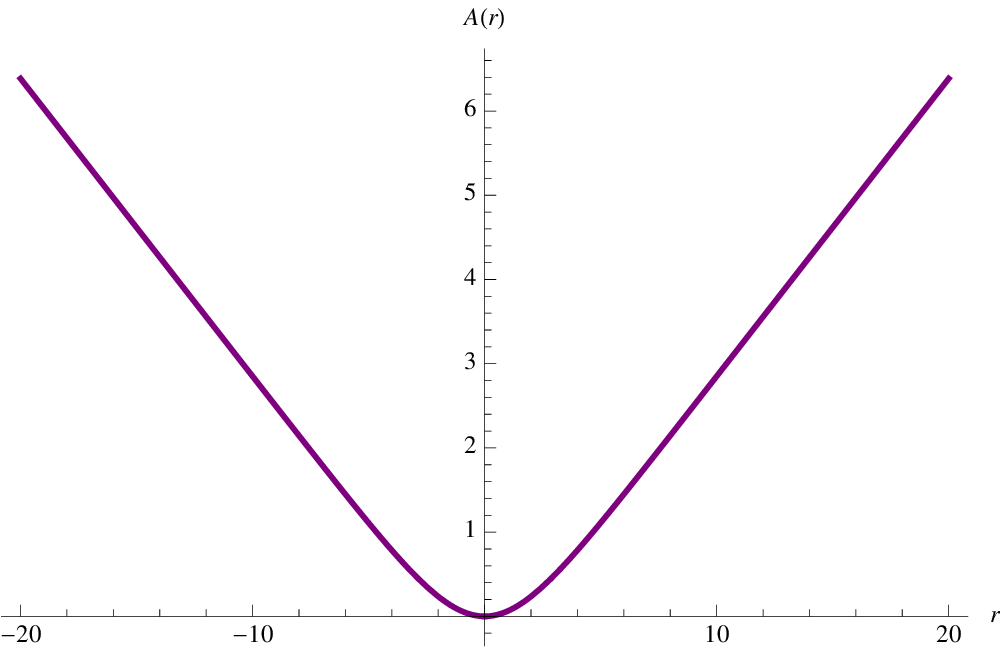}
                 \caption{Solution for $A$}
         \end{subfigure}
         \caption{$N=4$ Janus solution from type IIB compactification within a truncation to $\varphi_g$ and $\chi_g$ with $\kappa=\lambda=1$ and $\ell =2\sqrt{2}$}\label{fig3}
 \end{figure}

We first consider the Janus solution with only the dilaton and axion in the gravity multiplet non-vanishing. In this case, the BPS conditions $\delta\lambda^i_a=0$ are automatically satisfied by setting $\varphi_{1,2}=\chi_{1,2}=0$. By using the phase \eqref{Janus_phase} in $\delta\chi^i=0$ equations and separating real and imaginary parts, we obtain the following BPS equations
\begin{eqnarray}
\varphi_g'&=&-4\frac{A'}{W}\frac{\pd W}{\pd \varphi_g}-4\kappa e^{-\varphi_g}\frac{e^{-A}}{\ell W}\frac{\pd W}{\pd\chi_g},\nonumber \\
&=&\frac{-2\ell A'(e^{2\varphi_g}-1+2\chi_g^2e^{2\varphi_g})-4\kappa e^{\varphi_g-A}\chi_g}{\ell \left[(1+e^{\varphi_g})^2+\chi_g^2e^{2\varphi_g}\right]},\\
\chi_g'&=&-4\frac{A'}{W}e^{-2\varphi_g}\frac{\pd W}{\pd \chi_g}+4\kappa e^{-\varphi_g}\frac{e^{-A}}{\ell W}\frac{\pd W}{\pd \varphi_g},\nonumber \\
&=&\frac{2\kappa e^{-A-\varphi_g}(e^{2\varphi_g}-1+\chi_g^2e^{2\varphi_g})-4\ell \chi_gA'}{\ell \left[(1+e^{\varphi_g})^2+\chi_g^2e^{2\varphi_g}\right]},\\
0&=&A'^2+\frac{e^{-2A}}{\ell^2}-\frac{\lambda^2}{32}e^{-\varphi_g}\left[(1+e^{\varphi_g})^2+\chi_g^2e^{2\varphi_g}\right]
\end{eqnarray}
where we have also included the gravitini equations from \eqref{Janus_gravitino}. In term of the superpotential
\begin{equation}
W= \frac{\lambda}{4\sqrt{2}}e^{-\frac{\varphi_g}{2}}\sqrt{(1+e^{\varphi_g})^2+\chi_g^2e^{2\varphi_g}},
\end{equation}
these equations take a similar form as in the other four-dimensional Janus solutions in \cite{tri-sasakian-flow,warner_Janus,N3_Janus}. These equations solve all the BPS conditions for any $\epsilon^i$, $i=1,2,3,4$. Therefore, any solutions to these equations will preserve $N=4$ supersymmetry. We solve these equations numerically with an example of the solutions given in figure \ref{fig3}.\\
 \indent From figure \ref{fig3}, we see that the solution interpolates between $N=4$ $AdS_4$ vacua at both $r\rightarrow \pm \infty$. This solution is then interpreted as a $(1+1)$-dimensional conformal interface within the $N=4$ SCFT. The interface preserves $N=(4,0)$ supersymmetry on the interface due to the sign choice $\kappa=1$, and $SO(3)\times SO(3)$ symmetry remains unbroken throughout the solution.

\subsection{$N=1$ Janus solution}
\begin{figure}
         \centering
         \begin{subfigure}[b]{0.3\textwidth}
                 \includegraphics[width=\textwidth]{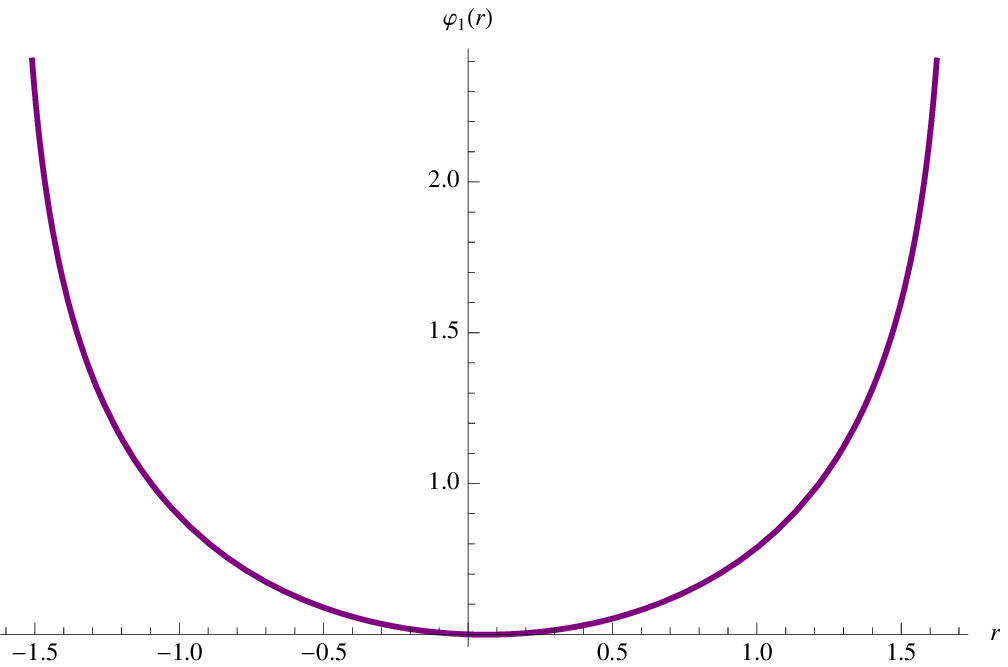}
                 \caption{Solution for $\varphi_1$}
         \end{subfigure}
\begin{subfigure}[b]{0.3\textwidth}
                 \includegraphics[width=\textwidth]{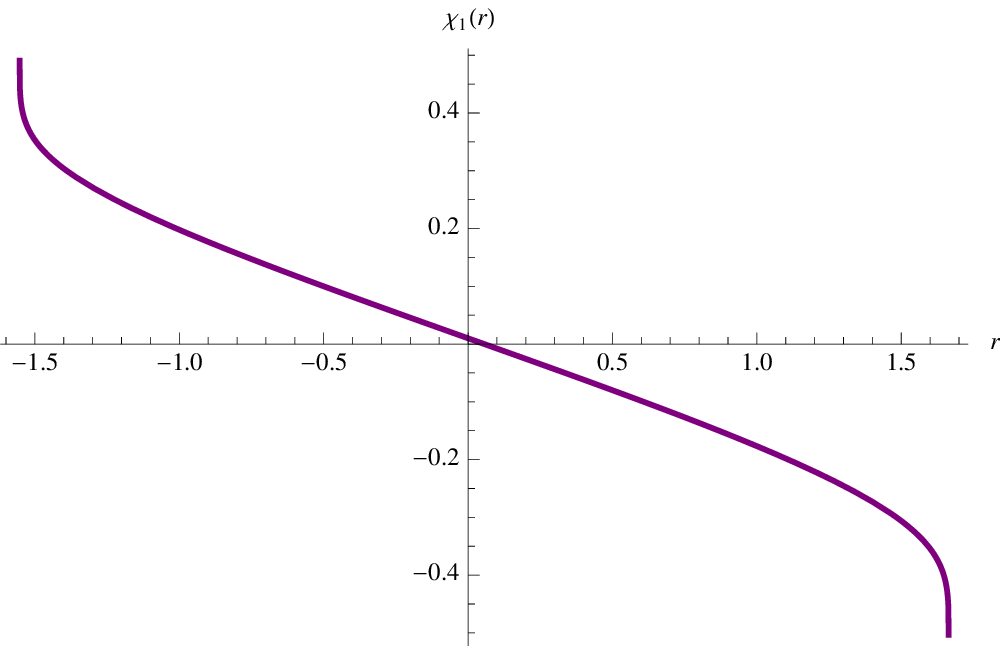}
                 \caption{Solution for $\chi_1$}
         \end{subfigure}%
         ~ %add desired spacing between images, e. g. ~, \quad, \qquad, \hfill etc.
         \begin{subfigure}[b]{0.3\textwidth}
                 \includegraphics[width=\textwidth]{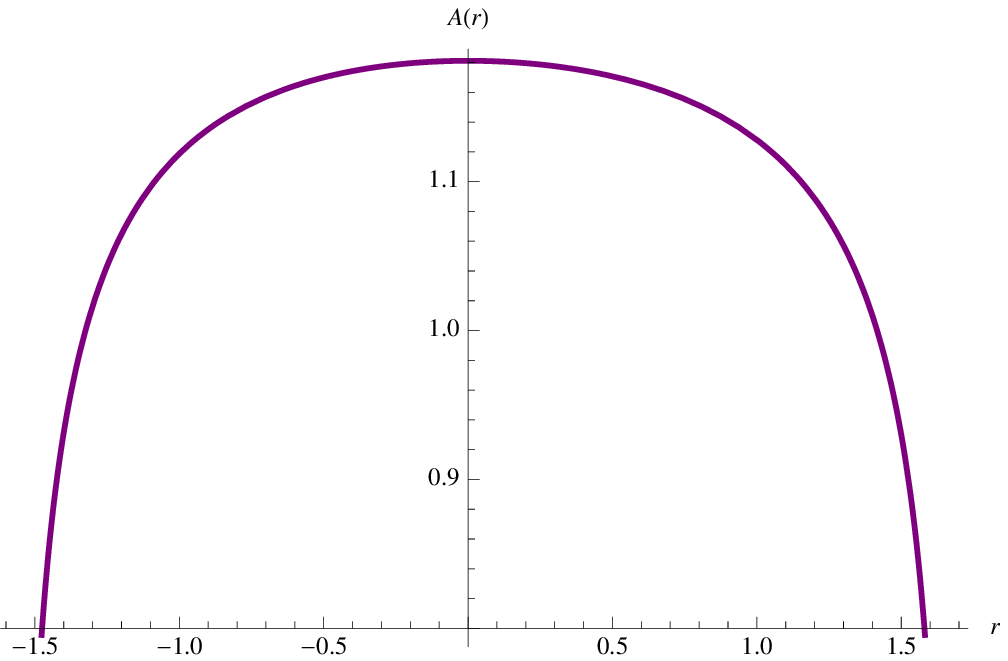}
                 \caption{Solution for $A$}
         \end{subfigure}
         \caption{$N=1$ Janus solution from type IIB compactification within a truncation to $\varphi_1$ and $\chi_1$ with $\kappa=\lambda=\ell=1$}\label{fig4}
 \end{figure}

The truncation keeping only $\varphi_1$ and $\chi_1$ is still consistent in the case of Janus BPS equations. In contrast to the previous truncation, any solutions to these equations will break $N=4$ supersymmetry to $N=1$ and preserve only $SO(3)$ diagonal subgroup of the full $SO(3)\times SO(3)$ symmetry of the $N=4$ $AdS_4$ vacuum.
\\
\indent The real superpotential for this truncation is given by
\begin{equation}
W=\frac{\lambda}{4\sqrt{2}}e^{\frac{\varphi_1}{2}}\sqrt{(e^{\varphi_1}-3)^2+9\chi_1^2e^{2\varphi_1}}
\end{equation}
in term of which the BPS equations can be written as
\begin{eqnarray}
\varphi_1'&=&-\frac{4}{3}\frac{A'}{W}\frac{\pd W}{\pd \varphi_1}-\frac{4}{3}\kappa e^{-\varphi_1}\frac{e^{-A}}{\ell W}\frac{\pd W}{\pd\chi_1},\nonumber \\
&=&\frac{2\ell A'(4e^{2\varphi_1}-3-9\chi_1^2e^{2\varphi_1}-e^{2\varphi_1})-12\kappa e^{\varphi_1-A}\chi_1}{\ell \left[(e^{\varphi_1}-3)^2+9\chi_1^2e^{2\varphi_1}\right]},\\
\chi_1'&=&-\frac{4}{3}\frac{A'}{W}e^{-2\varphi_1}\frac{\pd W}{\pd \chi_1}+\frac{4}{3}\kappa e^{-\varphi_1}\frac{e^{-A}}{\ell W}\frac{\pd W}{\pd \varphi_1},\nonumber \\
&=&\frac{2\kappa e^{-A-\varphi_1}(3-4e^{\varphi_1}+e^{2\varphi_1}+9\chi_1^2e^{2\varphi_1})-12\ell \chi_1A'}{\ell \left[(e^{\varphi_1}-3)^2+9\chi_1^2e^{2\varphi_1}\right]},\\
0&=&A'^2+\frac{e^{-2A}}{\ell^2}-\frac{\lambda^2}{32}e^{\varphi_1}\left[(e^{\varphi_1}-3)^2
+9\chi_1^2e^{2\varphi_1}\right].
\end{eqnarray}
Unlike the previous case, an intensive numerical search has not found any solutions interpolating between $AdS_4$ vacua in the limits $r\rightarrow \pm \infty$. All of the solutions found here are singular Janus in the sense that they connect singular domain walls at two finite values of the radial coordinate. We give an example of these solutions in figure \ref{fig4}.\\
\indent This solution could be interpreted as a conformal interface between two $N=1$ non-conformal phases of the dual $N=4$ SCFT. However, the singularities are of the bad type. An uplift to type IIB theory would be needed in order to decide whether the solution is physically acceptable in the ten-dimensional context.

%%%%%%%%%%%%%%%%%%%%%%%%%%%%%%%%%%%%%%%%%%%%%%%%%%%%%%%%%%%%%%%%%%%%%%%%%%%%%%%%%%%%%%%%%%%%%%%%%%%%%%%%%%%%%%%%%%%%%%%%%%%%%%%%%%%%%%%%%
\section{RG flows from type IIA geometric compactification}\label{IIA_flow}
\begin{figure}
         \centering
         \begin{subfigure}[b]{0.3\textwidth}
                 \includegraphics[width=\textwidth]{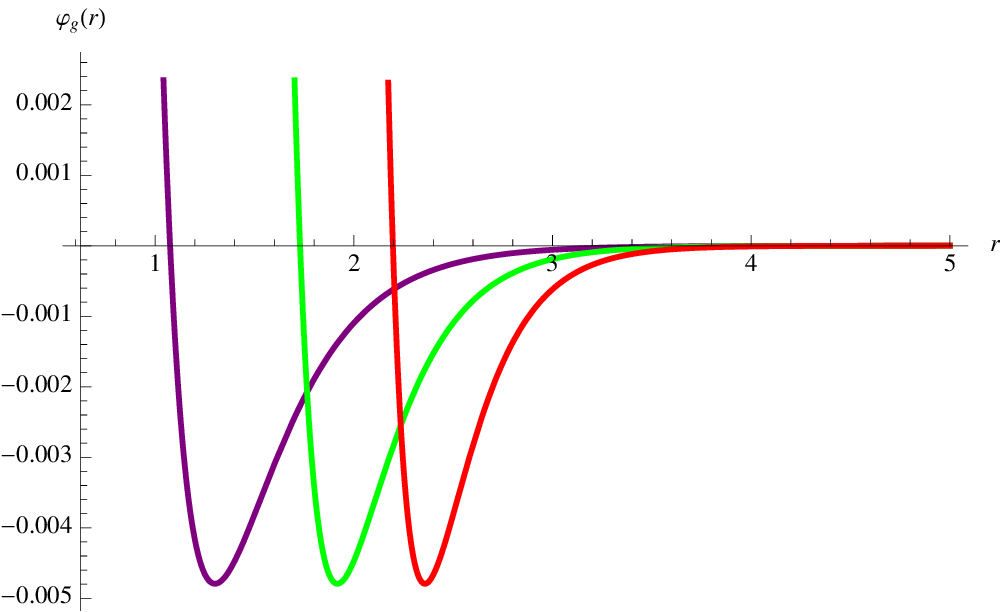}
                 \caption{Solution for $\varphi_g$}
         \end{subfigure}%
         ~ %add desired spacing between images, e. g. ~, \quad, \qquad, \hfill etc.
         \begin{subfigure}[b]{0.3\textwidth}
                 \includegraphics[width=\textwidth]{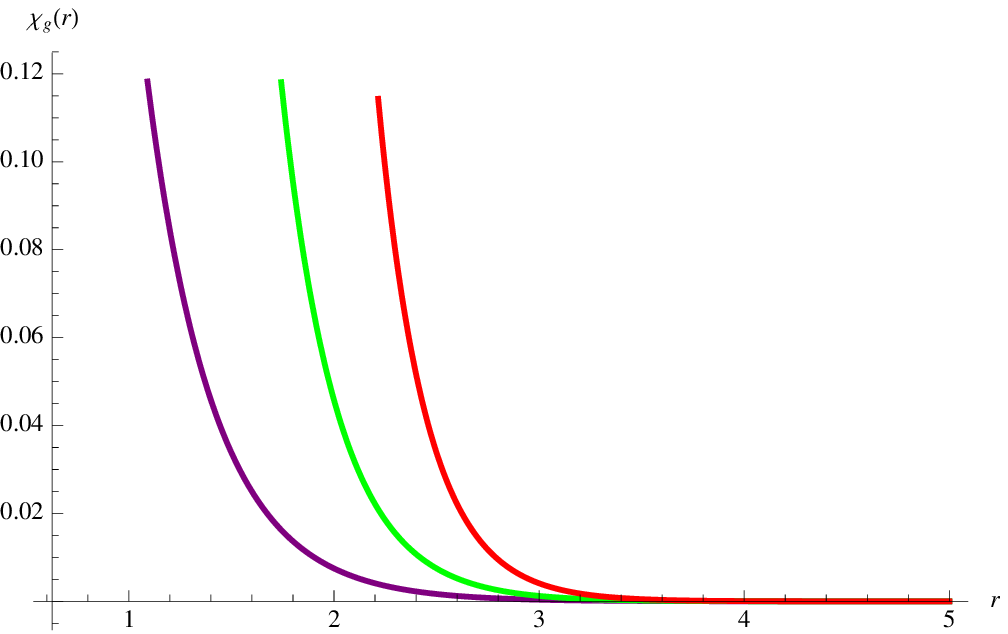}
                 \caption{Solution for $\chi_g$}
         \end{subfigure}
         \begin{subfigure}[b]{0.3\textwidth}
                 \includegraphics[width=\textwidth]{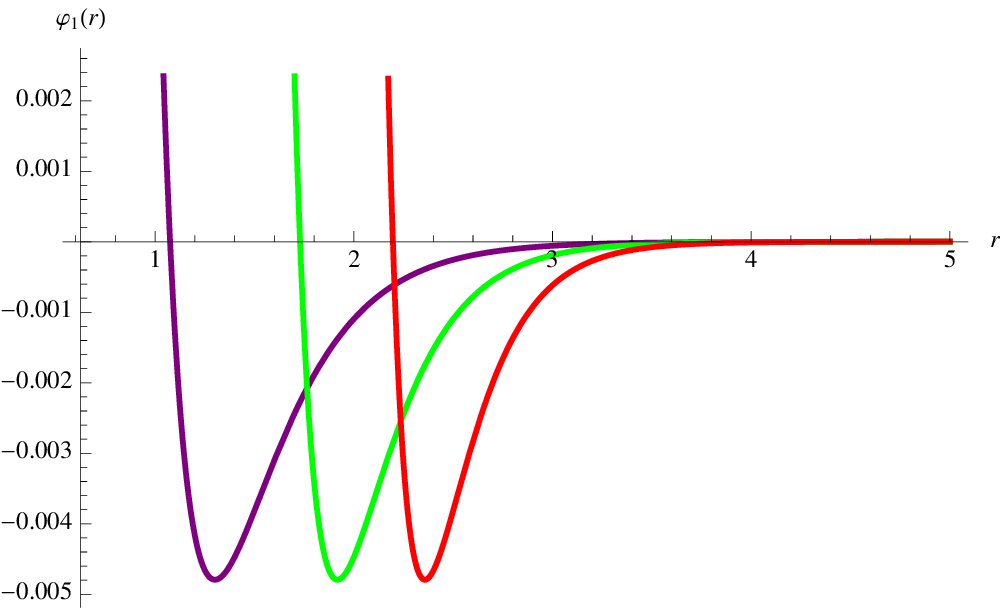}
                 \caption{Solution for $\varphi_1$}
         \end{subfigure}\\
\begin{subfigure}[b]{0.3\textwidth}
                 \includegraphics[width=\textwidth]{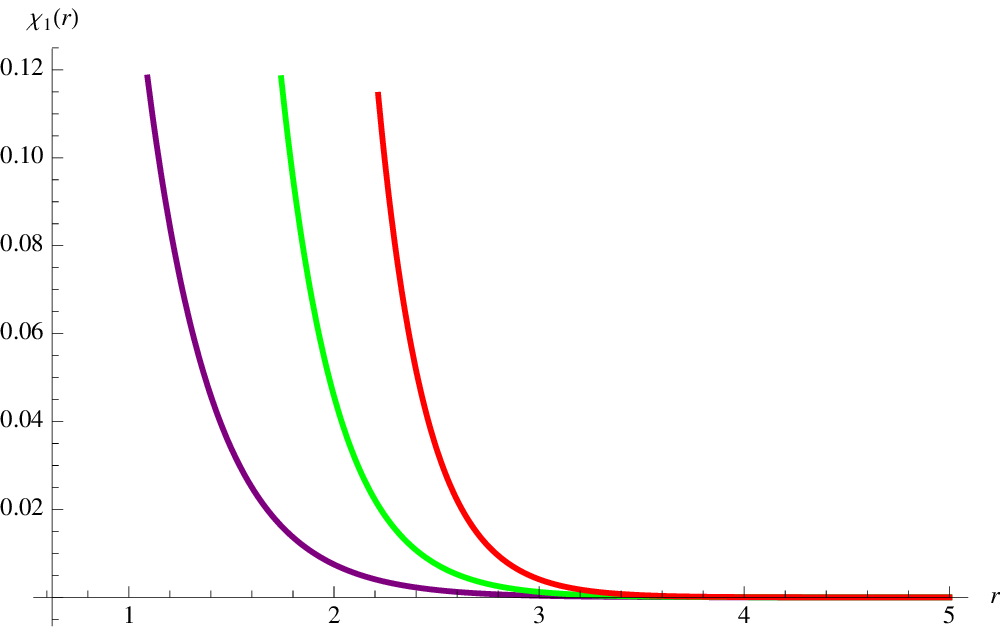}
                 \caption{Solution for $\chi_1$}
         \end{subfigure}%
         ~ %add desired spacing between images, e. g. ~, \quad, \qquad, \hfill etc.
         \begin{subfigure}[b]{0.3\textwidth}
                 \includegraphics[width=\textwidth]{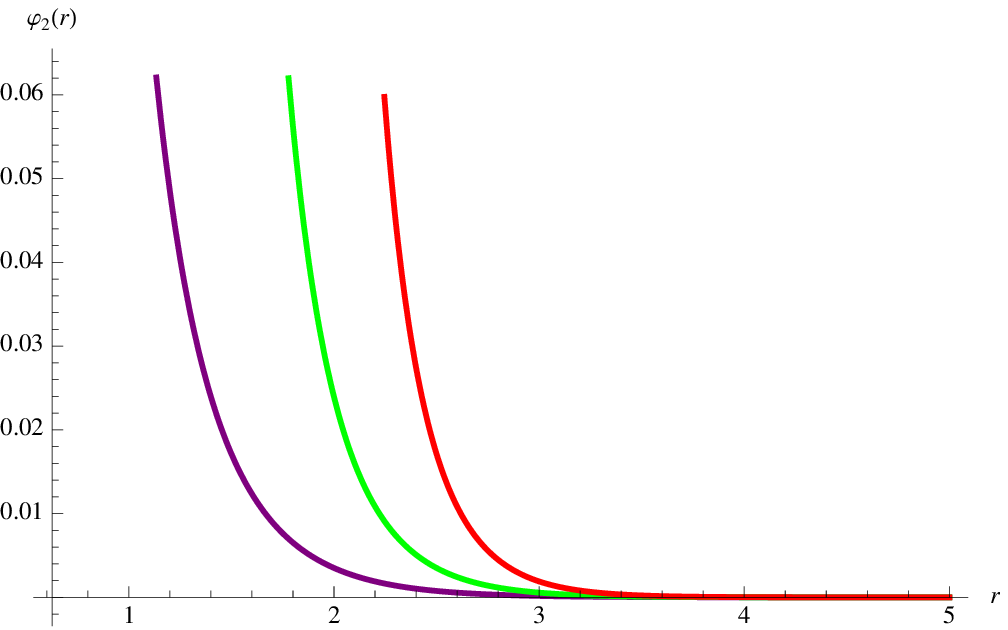}
                 \caption{Solution for $\varphi_2$}
         \end{subfigure}
         \begin{subfigure}[b]{0.3\textwidth}
                 \includegraphics[width=\textwidth]{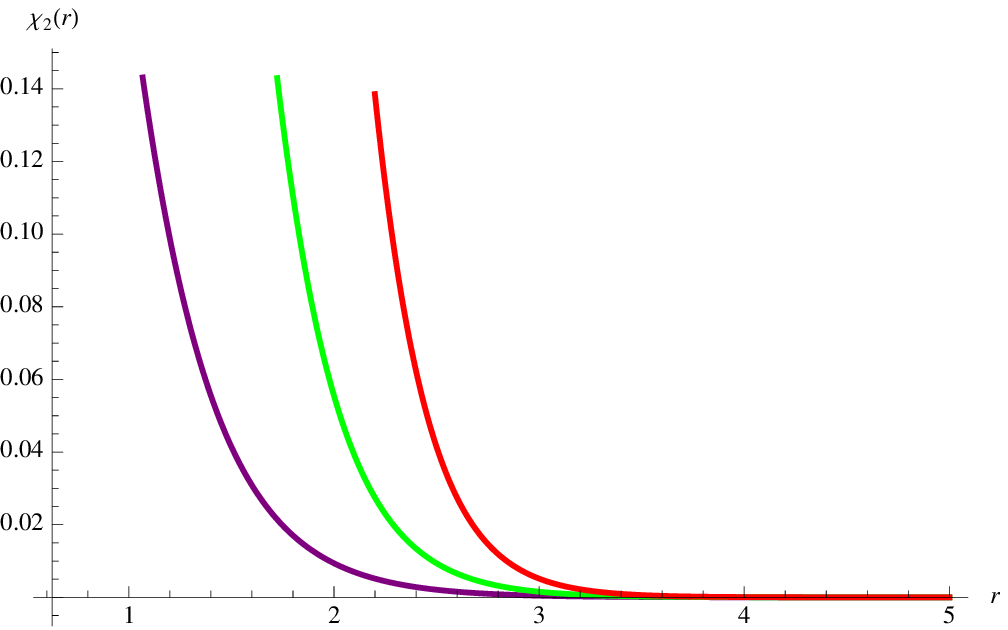}
                 \caption{Solution for $\chi_2$}
         \end{subfigure}\\
         \begin{subfigure}[b]{0.3\textwidth}
                 \includegraphics[width=\textwidth]{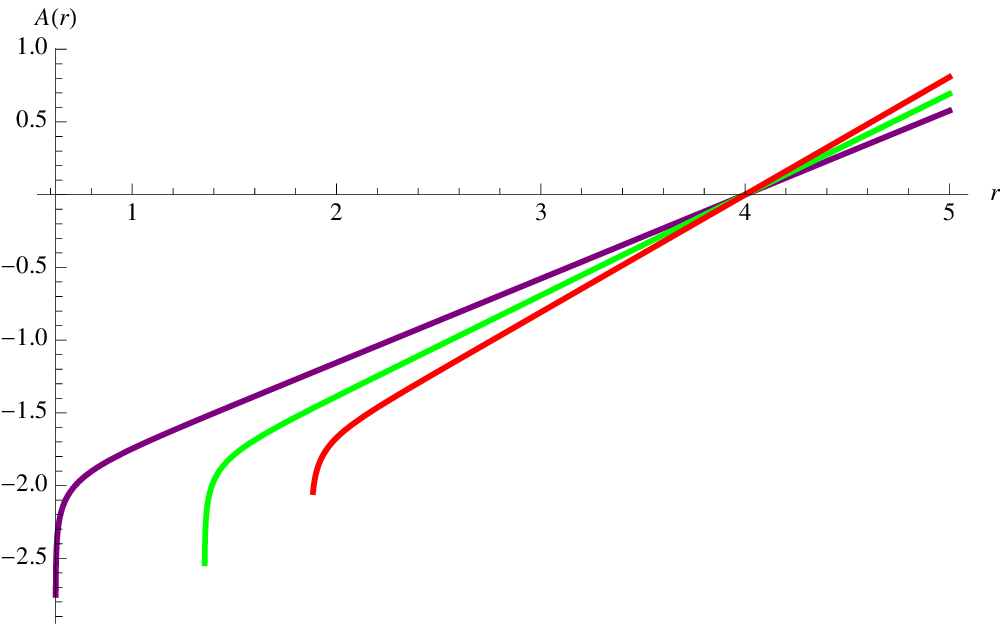}
                 \caption{Solution for $A$}
         \end{subfigure}
         \caption{$N=1$ RG flow solutions from type IIA compactification with $\lambda=1$ (purple), $\lambda=1.2$ (green) and $\lambda=1.4$ (red)}\label{fig5}
 \end{figure}

We now carry out a similar analysis for a geometric compactification of type IIA theory. The procedure is essentially the same, so we will omit unnecessary details. In this case, the compactification only involves gauge $(H_3,F_0,F_2,F_4,F_6)$ and geometric $(\omega)$ fluxes. However, the fluxes are more complicated and lead to many components of the embedding tensor compared to the type IIB case
\begin{eqnarray}
H_{ijk}&\sim &f_{-\bar{a}\bar{b}\bar{c}}=\Lambda_{-333}=\frac{\sqrt{6}}{3}\lambda,\qquad F_{aibjck}\sim f_{+\bar{a}\bar{b}\bar{c}}=\Lambda_{+333}=-\frac{3\sqrt{10}}{2}\lambda,\nonumber \\
F_{aibj}&\sim &f_{+\bar{a}\bar{b}\bar{k}}=\Lambda_{+334}=\frac{\sqrt{6}}{2}\lambda,\qquad F_{ai}\sim f_{+\bar{a}\bar{j}\bar{k}}=\Lambda_{+344}=\frac{\sqrt{10}}{6}\lambda,\nonumber \\
F_0&\sim &f_{+\bar{i}\bar{j}\bar{k}}=\Lambda_{+444}=\frac{5\sqrt{6}}{6}\lambda,\qquad H_{abk}\sim f_{+\bar{a}\bar{b}k}=\Lambda_{+233}=\frac{\sqrt{6}}{3}\lambda,\nonumber \\
{\omega_{ij}}^c&\sim &f_{-\bar{a}\bar{b}\bar{k}}=\Lambda_{-334}=\frac{\sqrt{10}}{3}\lambda,\nonumber \\
{\omega_{ka}}^j={\omega_{bk}}^i&=&{\omega_{bc}}^a\sim  f_{+\bar{a}\bar{j}k}=f_{+\bar{i}\bar{b}k}=f_{+a\bar{b}\bar{c}}=\Lambda_{+234}=\Lambda_{+133}=\sqrt{10}\lambda\, .
\end{eqnarray}
In the above equations, we have also given the form field corresponding to each flux component.
\\
\indent The resulting gauged $N=4$ supergravity has a non-semisimple group $ISO(3)\ltimes U(1)^6$ and admits the minimal $N=1$ $AdS_4$ vacuum at which the gauge group is broken down to $SO(3)$ compact subgroup. The corresponding superpotential for the unbroken $N=1$ supersymmetry is given by
\begin{eqnarray}
\mathcal{W} &=&\frac{\lambda}{24}e^{\frac{1}{2}(\varphi_1-3\varphi_2+\varphi_g)}\left[2e^{\varphi_1+2\varphi_2-\varphi_g}
\left[3\sqrt{5}i+e^{2\varphi_2}(\sqrt{3}+3\sqrt{5}\chi_2)\right](i+\chi_ge^{\varphi_g})\right.\nonumber \\
& &-5\sqrt{3}e^{\varphi_1}(i+e^{\varphi_2}\chi_2)^3-3\sqrt{5}e^{\varphi_1+\varphi_2}
(i+e^{\varphi_2}\chi_2)^2-9\sqrt{3}ie^{\varphi_1+2\varphi_2}\nonumber \\
& &+18\sqrt{5}e^{2\varphi_2}(i+e^{\varphi_1}\chi_1)(i+e^{\varphi_2}\chi_2)+6\sqrt{3}ie^{3\varphi_2}+9\sqrt{5}e^{\varphi_1+3\varphi_2}\nonumber \\
& &\left.+6\sqrt{3}e^{\varphi_1+3\varphi_2}\chi_1-9\sqrt{3}\chi_2e^{\varphi_1+3\varphi_2}\right].
\end{eqnarray}
The scalar potential can be written in term of $W=|\mc{W}|$ as
\begin{equation}
V=-\frac{1}{2}K^{ij}\frac{\pd W}{\pd \phi^i}\frac{\pd W}{\pd \phi^j}-\frac{3}{4}W^2\, .
\end{equation}
Its explicit form is given in the appendix.
\\
\indent When all scalars vanish, there is an $N=1$ $AdS_4$ vacuum with the cosmological constant
\begin{equation}
V_0=-\lambda^2\, .
\end{equation}
The six scalars have squared masses as follow
\begin{equation}
m^2L^2:\qquad 0,-2,4\pm \sqrt{6},\frac{1}{3}(47\pm \sqrt{159}).
\end{equation}
All of these values are in agreement with \cite{type_II_orbifold} after changing to our convention including a factor of $3$.
\\
\indent As in the type IIB case, the BPS equations obtained from supersymmetry variations can be written as
\begin{equation}
A'=W,\qquad {\varphi^i}'=K^{ij}\frac{\pd W}{\pd \phi^j}\, .
\end{equation}
However, the resulting equations are much more complicated than those from type IIB compactification. We will then not give them in this paper. Furthermore, we have not found any consistent subtruncation within this set of equations. In the following, we will only give examples of holographic RG flows from the $N=1$ SCFT dual to the above $AdS_4$ critical point to non-conformal $N=1$ field theories in the IR. These numerical solutions are shown in figure \ref{fig5} with three different values of the flux parameter $\lambda$ as in the IIB case. \\
\indent
As in the IIB case, we have numerically analyzed the scalar potential near the singularity and found that it leads to $V\rightarrow \infty$ which implies the singularity is unphysical.

%%%%%%%%%%%%%%%%%%%%%%%%%%%%%%%%%%%%%%%%%%%%%%%%%%%%%%%%%%%%%%%%%%%%%%%%%%%%%%%%%%%%%%%%%%%%%%%%%%%%%%%%%%%%%%%%%%%%%%%%%%%%%%%%%%%%%%%%%
\section{Conclusions and discussions}\label{conclusion}
We have found many supersymmetric RG flows and examples of Janus solutions in $N=4$ gauged supergravities obtained from flux compactifications of type II string theories. These solutions describe supersymmetric deformations and conformal interfaces within the dual $N=4$ and $N=1$ SCFTs in three dimensions. Many of the flow solutions have been obtained analytically which should be useful for further investigation.
\\
\indent In type IIB non-geometric compactification, the gauged supergravity has $ISO(3)\times ISO(3)$ gauge group and admits an $N=4$ $AdS_4$ vacuum dual to an $N=4$ SCFT with global symmetry $SO(3)\times SO(3)$. We have found two classes of supersymmetric RG flows. The first one preserves $N=4$ supersymmetry, and the global $SO(3)\times SO(3)$ symmetry is unbroken. This type of solutions can be obtained by turning on only the dilaton and axion in the gravity multiplet dual to relevant operators of dimensions $\Delta=1,2$. In this case, the flows are accordingly driven by relevant operators. Another possibility for preserving $N=4$ supersymmetry is to truncate out all axions or pseudoscalars. The resulting RG flows are driven by relevant and irrelevant operators of dimensions $\Delta=1,2$ and $\Delta=4$, respectively. When the axions in the vector multplets, corresponding to marginal deformations, are turned on, the flows break $N=4$ supersymmetry to $N=1$ and break $SO(3)\times SO(3)$ symmetry to their $SO(3)$ diagonal subgroup. We have given numerically the flows driven by marginal and irrelevant operators and the most general deformations in the presence of all types of deformations, relevant, marginal and irrelevant. It has been pointed out in \cite{type_II_orbifold} that the vacuum structure of type IIB compactification is very rich. The solutions found in this paper show that the number of supersymmetric deformations of these vacua is also enormous.
\\
\indent Within this type IIB compactification, we have also given Janus solutions preserving $N=4$ and $N=1$ supersymmetry. These correspond to $(1+1)$-dimensional conformal interfaces preserving $SO(3)\times SO(3)$ and $SO(3)$ symmetry. For the $N=4$ solution, we have given a numerical solution interpolating between $AdS_4$ vacua on the two sides of the interface, called regular Janus. This solution gives a holographic dual of a conformal interface in $N=4$ SCFT. For the $N=1$ case, we have not found this type of solutions but the singular Janus, interpolating between $N=1$ non-conformal phases of the dual $N=4$ SCFT. The situation is very similar to the $N=1$ Janus solutions studied in \cite{tri-sasakian-flow}. It would be interesting to have a definite conclusion about the existence of regular Janus solutions in these two $N=4$ gauged supergravities. 
\\
\indent In this non-geometric compactification, it is useful to give some comments about the holographic interpretation of the results. Due to its non-geometric nature, the stringy origin of the $N=4$ gauged supergravity is presently not well understood. This makes the meaning of the resulting solutions in terms of RG flows in the dual SCFT unclear. However, working in four-dimensional gauged supergravity has an obvious advantage in the sense that the whole formulation of N=4 gauged supergravity is virtually unchanged for all gaugings from both geometric and non-geometric compactifications. Therefore, the approach used here can be carried out for all other gaugings regardless of their higher dimensional origins. On the other hand, the full interpretation of the results in higher dimensional contexts calls for further study. Hopefully, the results presented here could be useful along this line of investigations. 
\\
\indent We have also carried out the same analysis in a geometric compactification of type IIA theory resulting in $N=4$ gauged supergravity with $ISO(3)\ltimes U(1)^6$. The gauged supergravity admits an $N=1$ $AdS_4$ vacuum dual to an $N=1$ SCFT in three dimensions. Due to the lack of further consistent subtruncation, we have numerically found examples of holographic RG flows to $N=1$ non-conformal field theories. Similar to the solutions in the type IIB case, the flows are driven by relevant, marginal and irrelevant operators in a more complicated manner. It should be pointed out that the massless scalars dual to marginal deformations considered in this paper are not the Goldstone bosons corresponding to the symmetry breaking $ISO(3)\times ISO(3)\rightarrow SO(3)\times SO(3)$ and $ISO(3)\ltimes U(1)^6\rightarrow SO(3)$. The Goldstone bosons transform non-trivially under the residual symmetry groups $SO(3)\times SO(3)$ and $SO(3)$ while the massless scalars considered in the solutions are singlets. Therefore, they are truly marginal deformations in the $N=1$ and $N=4$ SCFTs. Note also that, in the type IIB case, these marginal deformations break $N=4$ supersymmetry in consistent with the fact that all maximally supersymmetric $AdS_4$ vacua of $N=4$ gauged supergravity have no moduli preserving $N=4$ supersymmetry \cite{AdS4_N4_Jan}.
\\
\indent There are many possibilities for further investigations. First of all, it would be interesting to identify the $N=1$ and $N=4$ SCFTs dual to the $N=1$ and $N=4$ $AdS_4$ vacua. This should allow to identify the dual operators driving the RG flows obtained holographically in this paper. It could be interesting to look for more general Janus solutions in type IIB compactification with more scalars turned on and also look for similar solutions in type IIA compactification. Another direction would be to uplift the solutions found here to ten dimensions. This could be used to identify the $g_{00}$ component of the ten-dimensional metric and checked whether the unphysical singularities by the criterion of \cite{Gubser_singularity} are physical by the criterion of \cite{Maldacena_Nunez_nogo}. Finally, it would be of particular interest to further explore type IIB compactification with more general fluxes than those considered in \cite{type_II_orbifold}. This could enlarge the solution space of both $AdS_4$ vacua and their deformations including possible flow solutions between two $AdS_4$ vacua. We leave these issues for future works.
\vspace{0.5cm}\\
%%%%%%%%%%%%%%%%%%%%%%%%%%%%%%%%%%%%%%%%%%%%%%%%%%%%%%%%%%%%%%%%%%%%%%%%%%%%%%%%%%%%%%%%%%%%%%%%%%%%%%%%%%%%%%%%%%%%%%%%%%%%%%%%%%%%%%%%%
{\large{\textbf{Acknowledgement}}} \\
This work is supported by The Thailand Research Fund (TRF) under
grant RSA5980037.
%%%%%%%%%%%%%%%%%%%%%%%%%%%%%%%%%%%%%%%%%%%%%%%%%%%%%%%%%%%%%%%%%%%%%%%%%%%%%%%%%%%%%%%%%%%%%%%%%%%%%%%%%%%%%%%%%%%%%%%%%%%%%%%%%%%%%%%%%%%%%%%%%%%%%%%%%%%%%%%%
\appendix
\section{Useful formulae}
In this appendix, we collect all of the conventions about 't Hooft symbols, the scalar potential coming from type IIA geometric compactification and complicated BPS equations arising from type IIB non-geometric compactification with all $SO(3)$ singlet scalars non-vanishing.
\subsection{'t Hooft symbols}
To convert an $SO(6)$ vector index $m$ to a pair of anti-symmetric $SU(4)$ indices $[ij]$, we use the following 't Hooft symbols
\begin{align}
&G_{1}^{ij} = \left[
\begin{array}{cccc}
 0 & -1 & 0 & 0 \\
 1 & 0 & 0 & 0 \\
 0 & 0 & 0 & 1 \\
 0 & 0 & -1 & 0 \\
\end{array}
\right],
&G_{2}^{ij} = \left[
\begin{array}{cccc}
 0 & 0 & -1 & 0 \\
 0 & 0 & 0 & -1 \\
 1 & 0 & 0 & 0 \\
 0 & 1 & 0 & 0 \\
\end{array}
\right],\notag\\
&G_{3}^{ij} = \left[
\begin{array}{cccc}
 0 & 0 & 0 & -1 \\
 0 & 0 & 1 & 0 \\
 0 & -1 & 0 & 0 \\
 1 & 0 & 0 & 0 \\
\end{array}
\right],
&G_{4}^{ij} = \left[
\begin{array}{cccc}
 0 & -i & 0 & 0 \\
 i & 0 & 0 & 0 \\
 0 & 0 & 0 & -i \\
 0 & 0 & i & 0 \\
\end{array}
\right],\notag\\
&G_{5}^{ij} = \left[
\begin{array}{cccc}
 0 & 0 & -i & 0 \\
 0 & 0 & 0 & i \\
 i & 0 & 0 & 0 \\
 0 & -i & 0 & 0 \\
\end{array}
\right],
&G_{6}^{ij} = \left[
\begin{array}{cccc}
 0 & 0 & 0 & -i \\
 0 & 0 & -i & 0 \\
 0 & i & 0 & 0 \\
 i & 0 & 0 & 0 \\
\end{array}
\right].
\end{align}
These matrices satisfy the relation
\begin{equation}
G_{mij}=-\frac{1}{2}\epsilon_{ijkl}G^{kl}_{m}=-(G_{m}^{ij})^{*}\, .
\end{equation}

\subsection{BPS equations for type IIB compactification}
In this section, we give the full BPS equations for the non-geometric compactification of type IIB theory. These equations are given by
\begin{align}
 \varphi_{g}^{\prime} =& -\frac{1}{32W} e^{\varphi _1-3 \varphi _2-\varphi _g} \lambda ^2 [-e^{2 \varphi _1}-9 e^{2 \varphi _2}+6 e^{\varphi _1+\varphi _2}+e^{2   (\varphi _1+3 \varphi _2+\varphi _g)}\notag\\
 &+9 e^{4 \varphi _2+2 \varphi _g}-6 e^{\varphi _1+5 \varphi _2+2 \varphi _g}+6 e^{3 \varphi _2+2   \varphi _g} (2 e^{\varphi _1}-3 e^{\varphi _2}-e^{2 \varphi _1+\varphi _2}\notag\\
 &+2 e^{\varphi _1+2 \varphi _2}) \chi _2 \chi _g+2 e^{5 \varphi   _2+2 \varphi _g} (6 e^{\varphi _1}-9 e^{\varphi _2}+e^{2 \varphi _1+\varphi _2}) \chi _2^3 \chi _g+e^{2 (\varphi _1+\varphi   _g)} \chi _g^2\notag\\
 &+9 e^{2 (\varphi _2+\varphi _g)} \chi _g^2-6 e^{\varphi _1+\varphi _2+2 \varphi _g} \chi _g^2+3 e^{4 \varphi _2}   (e^{2 \varphi _1}+3 e^{2 \varphi _2}-2 e^{\varphi _1+\varphi _2}) \chi _2^4 (-1\notag\\
 &+e^{2 \varphi _g} \chi _g^2)+e^{2 \varphi _1+6   \varphi _2} \chi _2^6 (-1+e^{2 \varphi _g} \chi _g^2)+3 e^{2 \varphi _2} \chi _2^2 (-e^{2 \varphi _1}-6 e^{2 \varphi _2}+4   e^{\varphi _1+\varphi _2}\notag\\
 &+3 e^{4 \varphi _2+2 \varphi _g}+e^{2 \varphi _g} (e^{2 \varphi _1}+6 e^{2 \varphi _2}-4 e^{\varphi _1+\varphi   _2}) \chi _g^2)+9 e^{2 (\varphi _1+\varphi _2)} \chi _1^2 (1\notag\\
 &+e^{2 \varphi _2} \chi _2^2) (-1+e^{2   (\varphi _2+\varphi _g)}-2 e^{2 (\varphi _2+\varphi _g)} \chi _2 \chi _g+e^{2 \varphi _g} \chi _g^2+e^{2 \varphi _2} \chi   _2^2 (-1+e^{2 \varphi _g} \chi _g^2))\notag\\
 &-6 e^{2 (\varphi _1+\varphi _2)} \chi _1 (-e^{2 \varphi _g} (-1+e^{2   \varphi _2}) \chi _g+e^{4 \varphi _2+2 \varphi _g} \chi _2^2 \chi _g-e^{4 \varphi _2+2 \varphi _g} \chi _2^4 \chi _g\notag\\
 &-\chi _2 (1+e^{4   \varphi _2+2 \varphi _g}-e^{2 \varphi _g} \chi _g^2)+2 e^{2 \varphi _2} \chi _2^3 (-1+e^{2 \varphi _g} \chi _g^2)+e^{4 \varphi _2}   \chi _2^5 (-1+e^{2 \varphi _g} \chi _g^2))]
\end{align}

\begin{align}
 \chi_{g}^{\prime} =&-\frac{1}{16W} e^{\varphi _1-3 \varphi _2-\varphi _g} \lambda ^2 [3 e^{3 \varphi _2} (2 e^{\varphi _1}-3 e^{\varphi _2}-e^{2 \varphi   _1+\varphi _2}+2 e^{\varphi _1+2 \varphi _2}) \chi _2\notag\\
 &+e^{5 \varphi _2} (6 e^{\varphi _1}-9 e^{\varphi _2}+e^{2 \varphi _1+\varphi   _2}) \chi _2^3+(e^{\varphi _1}-3 e^{\varphi _2})^2 \chi _g+3 e^{2 \varphi _2} (e^{2 \varphi _1}+6 e^{2 \varphi _2}\notag\\
 &-4   e^{\varphi _1+\varphi _2}) \chi _2^2 \chi _g+3 e^{4 \varphi _2} (e^{2 \varphi _1}+3 e^{2 \varphi _2}-2 e^{\varphi _1+\varphi _2})   \chi _2^4 \chi _g+e^{2 \varphi _1+6 \varphi _2} \chi _2^6 \chi _g\notag\\
 &+9 e^{2 (\varphi _1+\varphi _2)} \chi _1^2 (1+e^{2 \varphi _2}   \chi _2^2) (-e^{2 \varphi _2} \chi _2+\chi _g+e^{2 \varphi _2} \chi _2^2 \chi _g)-3 e^{2 (\varphi _1+\varphi _2)} \chi   _1 (1\notag\\
 &-e^{2 \varphi _2}+e^{4 \varphi _2} \chi _2^2-e^{4 \varphi _2} \chi _2^4+2 \chi _2 \chi _g+4 e^{2 \varphi _2} \chi _2^3 \chi _g+2 e^{4   \varphi _2} \chi _2^5 \chi _g)]
\end{align}

\begin{align}
 \varphi_{1}^{\prime} =&-\frac{1}{32W} e^{\varphi _1-3 \varphi _2-\varphi _g} \lambda ^2 [e^{2 \varphi _1}+3 e^{2 \varphi _2}-4 e^{\varphi _1+\varphi _2}-4 e^{\varphi   _1+2 \varphi _2+\varphi _g}+6 e^{3 \varphi _2+\varphi _g}\notag\\
 &+e^{2 (\varphi _1+3 \varphi _2+\varphi _g)}+2 e^{2 \varphi _1+3 \varphi   _2+\varphi _g}-4 e^{\varphi _1+4 \varphi _2+\varphi _g}+3 e^{4 \varphi _2+2 \varphi _g}-4 e^{\varphi _1+5 \varphi _2+2 \varphi _g}\notag\\
 &+2 e^{3 \varphi   _2+2 \varphi _g} (4 e^{\varphi _1}-3 e^{\varphi _2}-3 e^{2 \varphi _1+\varphi _2}+4 e^{\varphi _1+2 \varphi _2}) \chi _2 \chi _g+2 e^{5   \varphi _2+2 \varphi _g} (4 e^{\varphi _1}\notag\\
 &-3 e^{\varphi _2}+e^{2 \varphi _1+\varphi _2}) \chi _2^3 \chi _g+e^{2 (\varphi   _1+\varphi _g)} \chi _g^2+3 e^{2 (\varphi _2+\varphi _g)} \chi _g^2-4 e^{\varphi _1+\varphi _2+2 \varphi _g} \chi _g^2\notag\\
 &+e^{2   \varphi _1+6 \varphi _2} \chi _2^6 (1+e^{2 \varphi _g} \chi _g^2)+e^{2 \varphi _2} \chi _2^2 (3 e^{2 \varphi _1}+6 e^{2 \varphi   _2}-8 e^{\varphi _1+\varphi _2}+6 e^{3 \varphi _2+\varphi _g}\notag\\
 &-6 e^{2 \varphi _1+3 \varphi _2+\varphi _g}+4 e^{\varphi _1+4 \varphi _2+\varphi   _g}+3 e^{4 \varphi _2+2 \varphi _g}+e^{2 \varphi _g} (3 e^{2 \varphi _1}+6 e^{2 \varphi _2}\notag\\
 &-8 e^{\varphi _1+\varphi _2}) \chi  _g^2)+e^{4 \varphi _2} \chi _2^4 (3 e^{2 \varphi _1}+3 e^{2 \varphi _2}-4 e^{\varphi _1+\varphi _2}+4 e^{\varphi _1+2 \varphi   _2+\varphi _g}\notag\\
 &+e^{2 \varphi _g} (3 e^{2 \varphi _1}+3 e^{2 \varphi _2}-4 e^{\varphi _1+\varphi _2}) \chi _g^2)+9 e^{2   (\varphi _1+\varphi _2)} \chi _1^2 (1+e^{2 \varphi _2} \chi _2^2) ((1+e^{\varphi _2+\varphi _g})^2\notag\\
 &-2 e^{2   (\varphi _2+\varphi _g)} \chi _2 \chi _g+e^{2 \varphi _g} \chi _g^2+e^{2 \varphi _2} \chi _2^2 (1+e^{2 \varphi _g} \chi   _g^2))-6 e^{2 (\varphi _1+\varphi _2)} \chi _1 (-e^{2 \varphi _g} (-1\notag\\
 &+e^{2 \varphi _2}) \chi _g+e^{4   \varphi _2+2 \varphi _g} \chi _2^2 \chi _g-e^{4 \varphi _2+2 \varphi _g} \chi _2^4 \chi _g+e^{4 \varphi _2} \chi _2^5 (1+e^{2 \varphi _g}   \chi _g^2)\notag\\
 &+2 e^{2 \varphi _2} \chi _2^3 (1+e^{\varphi _2+\varphi _g}+e^{2 \varphi _g} \chi _g^2)+\chi _2 (1+2 e^{\varphi   _2+\varphi _g}-2 e^{3 \varphi _2+\varphi _g}\notag\\
 &-e^{4 \varphi _2+2 \varphi _g}+e^{2 \varphi _g} \chi _g^2))]
\end{align}

\begin{align}
\varphi_{2}^{\prime} =&-\frac{1}{32W} e^{\varphi _1-3 \varphi _2-\varphi _g} \lambda ^2 [-e^{2 \varphi _1}-3 e^{2 \varphi _2}+4 e^{\varphi _1+\varphi _2}+2 e^{\varphi   _1+2 \varphi _2+\varphi _g}\notag\\
 &+e^{2 (\varphi _1+3 \varphi _2+\varphi _g)}-2 e^{\varphi _1+4 \varphi _2+\varphi _g}+3 e^{4 \varphi _2+2   \varphi _g}-4 e^{\varphi _1+5 \varphi _2+2 \varphi _g}+2 e^{4 \varphi _2+2 \varphi _g} (-3\notag\\
 &-e^{2 \varphi _1}+4 e^{\varphi _1+\varphi   _2}) \chi _2 \chi _g+2 e^{5 \varphi _2+2 \varphi _g} (4 e^{\varphi _1}-9 e^{\varphi _2}+e^{2 \varphi _1+\varphi _2}) \chi _2^3   \chi _g\notag\\
 &-e^{2 (\varphi _1+\varphi _g)} \chi _g^2-3 e^{2 (\varphi _2+\varphi _g)} \chi _g^2+4 e^{\varphi _1+\varphi _2+2   \varphi _g} \chi _g^2+e^{2 \varphi _1+6 \varphi _2} \chi _2^6 (1+e^{2 \varphi _g} \chi _g^2)\notag\\
 &+e^{2 \varphi _2} \chi _2^2 (-e^{2   \varphi _1}+6 e^{2 \varphi _2}+12 e^{3 \varphi _2+\varphi _g}-4 e^{2 \varphi _1+3 \varphi _2+\varphi _g}+6 e^{\varphi _1+4 \varphi _2+\varphi   _g}\notag\\
 &+9 e^{4 \varphi _2+2 \varphi _g}-e^{2 \varphi _g} (e^{2 \varphi _1}-6 e^{2 \varphi _2}) \chi _g^2)+e^{4 \varphi _2} \chi _2^4   (e^{2 \varphi _1}+9 e^{2 \varphi _2}-4 e^{\varphi _1+\varphi _2}\notag\\
 &+6 e^{\varphi _1+2 \varphi _2+\varphi _g}+e^{2 \varphi _g} (e^{2 \varphi   _1}+9 e^{2 \varphi _2}-4 e^{\varphi _1+\varphi _2}) \chi _g^2)-2 e^{2 (\varphi _1+\varphi _2)} \chi _1 (-e^{2 \varphi  _g} (1\notag\\
 &+e^{2 \varphi _2}) \chi _g+3 e^{4 \varphi _2+2 \varphi _g} \chi _2^2 \chi _g-3 e^{4 \varphi _2+2 \varphi _g} \chi _2^4 \chi _g+3   e^{4 \varphi _2} \chi _2^5 (1+e^{2 \varphi _g} \chi _g^2)\notag\\
 &+2 e^{2 \varphi _2} \chi _2^3 (1+2 e^{\varphi _2+\varphi _g}+e^{2 \varphi   _g} \chi _g^2)-\chi _2 (1+4 e^{3 \varphi _2+\varphi _g}+3 e^{4 \varphi _2+2 \varphi _g}+e^{2 \varphi _g} \chi _g^2))\notag\\
 &+3 e^{2   (\varphi _1+\varphi _2)} \chi _1^2 (-1+e^{2 (\varphi _2+\varphi _g)}-2 e^{2 (\varphi _2+\varphi _g)} \chi _2   \chi _g-6 e^{4 \varphi _2+2 \varphi _g} \chi _2^3 \chi _g-e^{2 \varphi _g} \chi _g^2\notag\\
 &+3 e^{4 \varphi _2} \chi _2^4 (1+e^{2 \varphi _g} \chi   _g^2)+e^{2 \varphi _2} \chi _2^2 (2+4 e^{\varphi _2+\varphi _g}+3 e^{2 (\varphi _2+\varphi _g)}+2 e^{2 \varphi _g} \chi   _g^2))]
\end{align}

\begin{align}
\chi_{1}^{\prime} =&\frac{1}{16W} e^{\varphi _1-\varphi _2-\varphi _g} \lambda ^2 [-e^{2 \varphi _g} (-1+e^{2 \varphi _2}) \chi _g+e^{4 \varphi _2+2   \varphi _g} \chi _2^2 \chi _g-e^{4 \varphi _2+2 \varphi _g} \chi _2^4 \chi _g\notag\\
 &+e^{4 \varphi _2} \chi _2^5 (1+e^{2 \varphi _g} \chi   _g^2)+2 e^{2 \varphi _2} \chi _2^3 (1+e^{\varphi _2+\varphi _g}+e^{2 \varphi _g} \chi _g^2)+\chi _2 (1+2 e^{\varphi   _2+\varphi _g}\notag\\
 &-2 e^{3 \varphi _2+\varphi _g}-e^{4 \varphi _2+2 \varphi _g}+e^{2 \varphi _g} \chi _g^2)-3 \chi _1 (1+e^{2 \varphi _2}   \chi _2^2) ((1+e^{\varphi _2+\varphi _g})^2\notag\\
 &-2 e^{2 (\varphi _2+\varphi _g)} \chi _2 \chi _g+e^{2 \varphi _g}   \chi _g^2+e^{2 \varphi _2} \chi _2^2 (1+e^{2 \varphi _g} \chi _g^2))]
\end{align}

\begin{align}
\chi_{2}^{\prime} =&-\frac{1}{16W} e^{\varphi _1-3 \varphi _2-\varphi _g} \lambda ^2 [e^{\varphi _2+2 \varphi _g} (2 e^{\varphi _1}-3 e^{\varphi _2}-e^{2   \varphi _1+\varphi _2}+2 e^{\varphi _1+2 \varphi _2}) \chi _g\notag\\
 &+e^{3 \varphi _2+2 \varphi _g} (6 e^{\varphi _1}-9 e^{\varphi _2}+e^{2   \varphi _1+\varphi _2}) \chi _2^2 \chi _g+e^{2 \varphi _1+4 \varphi _2} \chi _2^5 (1+e^{2 \varphi _g} \chi _g^2)\notag\\
 &+\chi _2   (e^{2 \varphi _1}+6 e^{2 \varphi _2}-4 e^{\varphi _1+\varphi _2}+6 e^{3 \varphi _2+\varphi _g}-2 e^{2 \varphi _1+3 \varphi _2+\varphi _g}+2   e^{\varphi _1+4 \varphi _2+\varphi _g}\notag\\
 &+3 e^{4 \varphi _2+2 \varphi _g}+e^{2 \varphi _g} (e^{2 \varphi _1}+6 e^{2 \varphi _2}-4 e^{\varphi   _1+\varphi _2}) \chi _g^2)+2 e^{2 \varphi _2} \chi _2^3 (e^{2 \varphi _1}+3 e^{2 \varphi _2}\notag\\
 &-2 e^{\varphi _1+\varphi _2}+2   e^{\varphi _1+2 \varphi _2+\varphi _g}+e^{2 \varphi _g} (e^{2 \varphi _1}+3 e^{2 \varphi _2}-2 e^{\varphi _1+\varphi _2}) \chi   _g^2)-e^{2 \varphi _1} \chi _1 (1\notag\\
 &+2 e^{\varphi _2+\varphi _g}-2 e^{3 \varphi _2+\varphi _g}-e^{4 \varphi _2+2 \varphi _g}+2 e^{4   \varphi _2+2 \varphi _g} \chi _2 \chi _g-4 e^{4 \varphi _2+2 \varphi _g} \chi _2^3 \chi _g\notag\\
 &+e^{2 \varphi _g} \chi _g^2+5 e^{4 \varphi _2} \chi _2^4   (1+e^{2 \varphi _g} \chi _g^2)+6 e^{2 \varphi _2} \chi _2^2 (1+e^{\varphi _2+\varphi _g}+e^{2 \varphi _g} \chi   _g^2))\notag\\
 &+3 e^{2 (\varphi _1+\varphi _2)} \chi _1^2 (-e^{2 \varphi _g} \chi _g-3 e^{2 (\varphi _2+\varphi _g)}   \chi _2^2 \chi _g+2 e^{2 \varphi _2} \chi _2^3 (1+e^{2 \varphi _g} \chi _g^2)\notag\\
 &+\chi _2 (2+2 e^{\varphi _2+\varphi _g}+e^{2   (\varphi _2+\varphi _g)}+2 e^{2 \varphi _g} \chi _g^2))]
\end{align}
where $W$ is given in \eqref{IIB_W}.

\subsection{Scalar potential from type IIA compactification}
The scalar potential obtained from a geometric compactification of type IIA theory is given by
\begin{align}
V=&\frac{1}{192} e^{\varphi _1-3 \varphi _2-\varphi _g} \lambda ^2 \left[20 e^{2 \varphi _1+4 \varphi _2}+25 e^{2 (\varphi   _1+\varphi _g)}-240 e^{\varphi _1+4 \varphi _2+\varphi _g}-180 e^{4 \varphi _2+2 \varphi _g}\notag\right.\\
   &+5 e^{2 (\varphi   _1+\varphi _2+\varphi _g)} (1+2 \sqrt{15} \chi _2+15 \chi _2^2)+12 e^{6 \varphi _2+2 \varphi _g} (1+2   \sqrt{15} \chi _2+15 \chi _2^2)\notag\\
   &+e^{2 \varphi _1+6 \varphi _2} (4+8 \sqrt{15} \chi _2+60 \chi _2^2)+e^{2   (\varphi _1+2 \varphi _2+\varphi _g)} [180 \chi _1^2-12 \chi _1 (3 \sqrt{15}\notag\\
   &+5 \chi _2   (2+\sqrt{15} \chi _2)-10 \chi _g)+3 (9+4 \sqrt{15} \chi _2-4 \sqrt{15} \chi _g)+5 [4   \sqrt{15} \chi _2^3\notag\\
   &+15 \chi _2^4-8 \chi _2 \chi _g+4 \chi _g^2+\chi _2^2 (22-4 \sqrt{15} \chi   _g)]]+e^{2 (\varphi _1+3 \varphi _2+\varphi _g)} \left[135\notag\right.\\
   &-54 \sqrt{15} \chi _2+10 \sqrt{15}   \chi _2^5+25 \chi _2^6+12 \sqrt{15} \chi _g-4 \chi _2^3 (3 \sqrt{15}+60 \chi _1+20 \chi _g)\notag\\
  &+8 \chi _2 (3   \chi _1+\chi _g) (18+3 \sqrt{15} \chi _1+\sqrt{15} \chi _g)-5 \chi _2^4 (-21+12 \sqrt{15} \chi _1\notag\\
   &+4   \sqrt{15} \chi _g)+4 [9 \chi _1 (\sqrt{15}+\chi _1)+6 \chi _1 \chi _g+\chi _g^2]+\chi _2^2   \left[-9-40 \sqrt{15} \chi _g\notag\right.\\
   &\left.\left.\left.+60 [9 \chi _1^2-2 \chi _1 (\sqrt{15}-3 \chi _g)+\chi   _g^2]\right]\right]\right].
\end{align}

%%%%%%%%%%%%%%%%%%%%%%%%%%%%%%%%%%%%%%%%%%%%%%%%%%%%%%%%%%%%%%%%%%%%%%%%%%%%%%%%%%%%%%%%%%%%%%%%%%%%%%%%%%%%%%%%%%%%%%%%%%%%%%%%%%%%%%%%%%%%%%%%%%%%%%%%%%%%%%%%

\end{document}